\documentclass[showpacs,floatfix,amssymb]{article}
\usepackage{graphicx,subfigure,amsmath,color,amsmath,amssymb,hyperref,cite}
\usepackage{hyperref, cite}
\begin{document}
\title{Investigating the hybrid textures of neutrino mass matrix for near maximal atmospheric neutrino mixing }

\author{Madan Singh$^{*} $\\
\it Department of Physics, National Institute of Technology Kurukshetra,\\
\it Haryana,136119, India.\\
\it $^{*}$singhmadan179@gmail.com
}
\maketitle

\begin{abstract}
In the present paper,  we have studied that the implication of a large value of the effective Majorana neutrino mass in  case of neutrino mass matrices having either two equal elements  and one zero element (popularly known as hybrid texture) or two equal cofactors and one zero minor (popularly known as inverse hybrid texture) in the flavor basis. In each of these cases, four out of sixty phenomenologically possible patterns predict near maximal atmospheric neutrino mixing angle in the limit of large effective Majorana neutrino mass. This feature remains irrespective of the experimental data on solar and reactor mixing angles. In addition, we have also performed the comparative study of all the viable cases of hybrid and inverse hybrid textures at 3$\sigma$ CL.
\end{abstract}

\section{Introduction}
In leptonic sector, the reactor mixing angle ($\theta_{13}$) has been established to a reasonably good degree of precision \cite{1, 2, 3, 4, 5}, and its non zero and relatively large value has not only provided an opportunity in exploring CP violation and the neutrino mass ordering in the future experiments, but has also highlighted the puzzle of neutrino mass and mixing pattern.  In spite of the significant developments made over the years, there are still several intriguing questions in the neutrino sector which remain unsettled. For instance, the present available data is unable to throw any light on the neutrino mass spectrum,
which may be normal/inverted and may even be degenerate. Another important issue is the determination of octant of atmospheric mixing angle $\theta_{23}$, which may be greater than or less than or equal to $45^{0}$.  The determination of the nature of neutrinos whether Dirac or Majorana also remains an open question. The  observation of neutrinoless double beta ($0\nu\beta\beta$) decay would eventually establish the Majorana nature of neutrinos.\\
The effective Majorana mass term related to $0\nu\beta\beta$ decay  can be expressed as
\begin{equation}
|M|_{ee}=|m_{1}c_{12}^{2}c_{13}^{2}e^{2i\rho}+m_{2}s_{12}^{2}c_{13}^{2}e^{2i\sigma}+m_{3}s_{13}^{2}|.
\end{equation}
Data from KamLAND-Zen experiment has  presented an improved search for neutrinoless double-beta ($0\nu\beta\beta$) decay  \cite{6} and it is found that  $|M|_{ee} <(0.061-0.165)eV$ at 90 $\%$ (or $<2\sigma$) CL. For recent reviews on $0\nu\beta\beta$ decay see Refs.\cite{7,8,9,10}.\\ 
 In the lack of any convincing theory, several phenomenological ideas have been proposed in the literature  so as to restrict the form of neutrino mass matrix, such as some elements of neutrino mass matrix are considered to be zero or equal \cite{11,12,13,14} or  some co-factors of neutrino mass matrix to be either zero or equal \cite {12,15,16,17}. Specifically,  mass matrices with zero textures (or cofactors) have been extensively studied \cite{11, 15} due to their connections to flavor symmetries. In addition, texture structures with one zero element (or minor) and an equality between two independent elements (or cofactors) in neutrino mass matrix have also been  studied in the literature \cite{13,14,17}. Such form of texture structures sets to one constraint equation and thus reduces the number of real free parameters of neutrino mass matrix to seven.  Hence they are considered as predictive as the well-known two-zero textures and can also be realised within the framework of seesaw mechanism.  Out of sixty possibilities,
only fifty four are found to be compatible with the neutrino
oscillation data \cite{14} for texture structures having
one zero element and an equal matrix elements in the
neutrino mass matrix (1TEE),
while for texture with one vanishing minor and an equal
cofactors in the neutrino mass matrix (1TEC) only fifty two cases are able to
survive the data \cite{17}. 
 
The purpose of present paper is to investigate the implication of large effective neutrino mass $|M|_{ee}$ on  1TEE and 1TEC structures of neutrino mass matrix, while taking into account the assumptions of Refs. \cite{18, 19}. The consideration of large $|M|_{ee}$ is motivated by the extensive search for this parameter in the ongoing $0\nu\beta\beta$ experiments.  The implication of large $|M|_{ee}$ has earlier been studied  for the viable cases of texture two zero and two vanishing minor, respectively \cite{18,19}.  Grimus et. al \cite{20} also predicted the near maximal atmospheric mixing for two zero textures when supplemented with the assumption of quasi degenerate mass spectrum. However the observation made in all these analyses are independent of solar and reactor mixing angles. Motivated by these works, we find that only four out of sixty cases are able to predict near maximal $\theta_{23}$ for 1TEE and 1TEC, respectively. In addition, the analysis also hints towards the indistinguisble feature of 1TEE and 1TEC.   To present the indistingusible nature of the 1TEE and 1TEC texture structures, we have then carried out a comparative study of all the viable cases of 1TEE and 1TEC at 3$\sigma$ CL.  The similarity between texture zero structures with one mass ordering and correponding cofactor zero structures with the opposite mass ordering has earlier been noted in Refs. \cite{21,22}.  In Ref.\cite{12}, the strong similarities have also been noted between the texture structures with two equalities of elements  and structures with two equalities of cofactors in neutrino mass matrix, with opposite mass ordering.

The rest of the paper is planned in following manner: In Section 2, we shall discuss the methodology to obtain the constraint equations. Section 3 is devoted to numerical analysis. Section 4 will summarize our result. 
 
\section{Methodology} 

The effective Majorana neutrino mass matrix $(M_{\nu})$ contains nine parameters which include three neutrino masses ($m_{1}$, $m_{2}$, $m_{3}$), three mixing angles ($\theta_{12}$, $\theta_{23}$, $\theta_{13}$) and three CP violating phases ($\delta$, $\rho$, $\sigma$). In the flavor basis, the Majorana neutrino mass matrix can be expressed as,
\begin{equation}
 M_{\nu}=P_{l}UP_{\nu}M^{\textrm{diag}}P_{\nu}^{T}U^{T}P_{l}^{T},
\end{equation}
where $M^{\textrm{diag}}$ = diag($m_{1}$, $m_{2}$, $m_{3}$) is the diagonal matrix of neutrino masses and $U$ is the flavor mixing matrix, 
and 
\begin{equation}
P_{\nu}=\left(
\begin{array}{ccc}
    e^{i\rho}& 0& 0 \\
  0 & e^{i\sigma} & 0\\
  0& 0&1  \\
\end{array}
\right), \qquad \qquad P_{l}= \left(
\begin{array}{ccc}
    e^{i \phi_{e}}& 0& 0 \\
  0 & e^{i \phi_{\mu}} & 0\\
  0& 0& e^{i \phi_{\tau}} \\
\end{array}
\right);
\end{equation}
where $P_{\nu}$ is diagonal phase matrix containing Majorana neutrinos $\rho, \sigma$. $P_{l}$ is unobservable phase matrix and depends on phase convention. 
Eq. (2) can be re-written as
\begin{equation}
M_{\nu}=P_{l}U\left(
\begin{array}{ccc}
    \lambda_{1}& 0& 0 \\
  0 & \lambda_{2} & 0\\
  0& 0& \lambda_{3} \\
\end{array}
\right)U^{T}P_{l}^{T},
\end{equation}
where
$\lambda_{1} = m_{1} e^{2i\rho},\lambda_{2} = m_{2} e^{2i\sigma} ,\lambda_{3} = m_{3}.$
For the present analysis, we consider the following parameterization of $U$ \cite{13}:
\begin{equation}
U=\left(
\begin{array}{ccc}
 c_{12}c_{13}& s_{12}c_{13}& s_{13} \\
-c_{12}s_{23}s_{13}-s_{12}c_{23}e^{-i\delta} & -s_{12}s_{23}s_{13}+c_{12}c_{23}e^{-i\delta} & s_{23}c_{13}\\
 -c_{12}c_{23}s_{13}+s_{12}s_{23}e^{-i\delta}& -s_{12}c_{23}s_{13}-c_{12}s_{23}e^{-i\delta}& c_{23}c_{13} \\
\end{array}
 \right),
\end{equation}
where, $c_{ij} = \cos \theta_{ij}$, $s_{ij}= \sin \theta_{ij}$. Here, $U$ is a 3 $\times$ 3 unitary matrix consisting of three flavor mixing angles ($\theta_{12}$, $\theta_{23}$, $\theta_{13}$) and one Dirac CP-violating phase $\delta$.\\

For hybrid texture structure (1TEE) of $M_{\nu}$, we can express the ratios
of neutrino mass eigenvalues in terms of the mixing matrix elements as  \cite{14}
\begin{equation}
\dfrac{\lambda_{1}}{\lambda_{3}} =\frac{P(U_{a3}U_{b3}U_{\alpha 2}U_{\beta 2}-U_{a2}U_{b2}U_{\alpha 3}U_{\beta 3})+(U_{a2}U_{b2}U_{c3}U_{d3}-U_{a3}U_{b3}U_{c2}U_{d2})} {P(U_{a2}U_{b2}U_{\alpha 1}U_{\beta 1}-U_{a1}U_{b1}U_{\alpha 2}U_{\beta 2})+(U_{a1}U_{b1}U_{c2}U_{d2}-U_{a2}U_{b2}U_{c1}U_{d1})}, 
\end{equation} 
\begin{equation}
\dfrac{\lambda_{2}}{\lambda_{3}} =\frac{P(U_{a1}U_{b1}U_{\alpha 3}U_{\beta 3}-U_{a3}U_{b3}U_{\alpha 1}U_{\beta 1})+(U_{a3}U_{b3}U_{c1}U_{d1}-U_{a1}U_{b1}U_{c3}U_{d3})} {P(U_{a2}U_{b2}U_{\alpha 1}U_{\beta 1}-U_{a1}U_{b1}U_{\alpha 2}U_{\beta 2})+(U_{a1}U_{b1}U_{c2}U_{d2}-U_{a2}U_{b2}U_{c1}U_{d1})},
\end{equation}
where $P= e^{i(\phi_{\alpha}+\phi_{\beta}-\phi_{c}-\phi_{d})}$ is a phase factor.
Similarly, in case of inverse hybrid texture structure (1TEC) of $M_{\nu}$ , we can express the ratios
of mass eigenvalues as \cite{17} follows
\begin{equation}
 \dfrac{\lambda_{1}}{\lambda_{3}} =\frac{A_{1}B_{2}-A_{2}B_{1}}{A_{2}B_{3}-A_{3}B_{2}},
 \end{equation}
 \begin{equation}
 \dfrac{\lambda_{2}}{\lambda_{3}} =\frac{A_{1}B_{2}-A_{2}B_{1}}{A_{3}B_{1}-A_{1}B_{3}},
 \end{equation} 
 where 
 \begin{equation}
 A_{i}=(U_{pj}U_{qj}U_{rk}U_{sk}-U_{tj}U_{uj}U_{vk}U_{wk})+(j \leftrightarrow k)
 \end{equation}
 \begin{eqnarray}
 &&\quad\qquad  B_{i}=(-1)^{m+n}Q(U_{aj}U_{bj}U_{ck}U_{dk}-U_{ej}U_{fj}U_{gk}U_{hk}),
 \nonumber \\
 &&~~~~~~-(-1)^{m^{'}+n^{'}}(U_{a^{'}j}U_{b^{'}j}U_{c^{'}k}U_{d^{'}k}-U_{e^{'}j}U_{f^{'}j}U_{g^{'}k}U_{h^{'}k})+(j \leftrightarrow k),
 \end{eqnarray}
 
 with ($i, j, k$) a cyclic permutation of (1, 2, 3) and $Q = e^{i(\phi_{a}+\phi_{b}+\phi_{c}+\phi_{d}-\phi_{a^{'}}-\phi_{b^{'}}-\phi_{c^{'}}-\phi_{d^{'}})}$ is phase factor.

 Using above expressions, we can obtain the magnitude of neutrino mass ratios, $\alpha\equiv\frac{|\lambda_{1}|}{|\lambda_{3}|} $ and $ \beta \equiv \frac{|\lambda_{2}|}{|\lambda_{3}|} $ in each texture structure, and the Majorana phases ($\rho, \sigma$) can be given as $\rho=\frac{1}{2}arg \left(\frac{\lambda _{1}}{\lambda _{3}}\right)$ and
$\sigma=\frac{1}{2}arg\left(\frac{\lambda _{2}}{\lambda _{3}}\right)$.
 
 The solar and atmospheric mass squared differences ($\delta m^{2}, \Delta m^{2}$), where $\delta m^{2}$ corresponds to solar mass-squared difference and $\Delta m^{2}$ corresponds to atmospheric mass-squared difference, can be defined as \cite{13}
\begin{equation}
 \delta m^{2}=(m_{2}^{2}-m_{1}^{2}),\;
 \end{equation}
 \begin{equation}
  \Delta m^{2}=m_{3}^{2}-\frac{1}{2}(m_{1}^{2}+m_{2}^{2}).
 \end{equation}
  The experimentally determined solar and atmospheric neutrino mass-squared differences can be related to neutrino mass ratios $(\alpha, \beta$) as
 \begin{equation}
 R_{\nu} \equiv\frac{\delta m^{2}} {|\Delta m^{2}|} =\frac{2(\beta ^{2}-\alpha ^{2})}{\left |2-(\beta ^{2}+\alpha ^{2})  \right |},
 \end{equation}
 and the three neutrino masses  can be determined in terms of $\alpha, \beta$ as
 \begin{equation}
 m_{3}=\sqrt{\dfrac{\delta m^{2}}{\beta^{2}-\alpha^{2}}},  \qquad m_{2}=m_{3} \beta, \qquad m_{1}=m_{3} \alpha.
 \end{equation}

 Among the sixty logically possible cases of 1TEE or 1TEC texture structures, there are certain pair, which exhibits similar phenomenological implications and are related via permutation symmetry \cite{14,17}.  This corresponds to permutation of the 2-3 rows and 2-3 columns of $M_{\nu}$. The corresponding permutation matrix can be given by
\begin{equation}
 P_{23} = \left(
\begin{array}{ccc}
    1& 0& 0 \\
  0 & 0 & 1\\
  0& 1& 0 \\
\end{array}
\right).
\end{equation}
With the help of permutation symmetry, one obtains the following relations among the neutrino oscillation parameters
\begin{equation}
\theta_{12}^{X}=\theta_{12}^{Y}, \ \
\theta_{23}^{X}=90^{\circ}-\theta_{23}^{Y},\ \
\theta_{13}^{X}=\theta_{13}^{Y}, \ \ \delta^{X}=\delta^{Y} -180^{\circ},
\end{equation}
where X and Y denote the cases related by 2-3 permutation. The following pair among sixty cases are related via permutation symmetry:
 \\
($A_{1}, A_{1});\quad (A_{2}, A_{8});\quad (A_{3}, A_{7});\quad (A_{4}, A_{6});\quad (A_{5}, A_{5});\quad (A_{9}, A_{10});\quad (B_{1}, C_{1}); \\
    \qquad         (B_{2}, C_{7}); \quad(B_{3}, C_{6}); \quad(B_{4},C_{5});\quad (B_{5}, C_{4}); \quad (B_{6}, C_{3})\quad(B_{7}, C_{2}); \quad(B_{8}, C_{10}); \\
   \qquad        (B_{9}, C_{9});\quad (B_{10}, C_{8}); \quad(D_{1}, F_{2}); \quad (D_{2}, F_{1});\quad (D_{3}, F_{4});\quad(D_{4}, F_{3}); \\
    \qquad       (D_{5}, F_{5});\quad (D_{6}, F_{9});\quad (D_{7}, F_{8});\quad (D_{8}, F_{7}); \quad(D_{9}, F_{6});\quad (D_{10}, F_{10});\\
    \qquad        (E_{1}, E_{2});\quad (E_{3}, E_{4});\quad (E_{5}, E_{5});\quad(E_{6}, E_{9});\quad (E_{7}, E_{8});\quad (E_{10}, E_{10}).$
        \\
      Clearly we are left with only thirty two independent cases. It is worthwhile to mention that cases $A_{1}, A_{5}, E_{5}$ and $ E_{10}$ are invariant under the permutations of 2- and 3-rows and columns.
\begin{table}
\begin{small}
\begin{center}
\begin{tabular}{|c|c|c|}
  \hline
  Parameter& Best Fit & 3$\sigma$ \\
  \hline
   $\delta m^{2}$ $[10^{-5}eV^{2}]$ & $7.50$& $7.03$ - $8.09$  \\
   \hline
   $|\Delta m^{2}_{31}|$ $[10^{-3}eV^{2}]$ (NO) & $2.52$ & $2.407$ - $2.643$ \\
   \hline
  $|\Delta m^{2}_{31}|$ $[10^{-3}eV^{2}]$ (IO) & $2.52$ &  $2.39$ - $2.63$ \\
  \hline
  $\theta_{12}$ & $33.56^{\circ}$ & $31.3^{\circ}$ - $35.99^{\circ}$\\
  \hline
  $ \theta_{23}$ (NO) & $41.6^{\circ}$  & $38.4^{\circ}$ - $52.8^{\circ}$ \\
  \hline
  $\theta_{23}$ (IO)& $50.0^{\circ}$ &  $38.8^{\circ}$ - $53.1^{\circ}$ \\
  \hline
  $\theta_{13}$ (NO) & $8.46^{\circ}$ &  $7.99^{\circ}$ - $8.90^{\circ}$ \\
  \hline
  $\theta_{13}$ (IO) & $8.49^{\circ}$ &  $8.03^{\circ}$ - $8.93^{\circ}$ \\
  \hline
  $\delta$ (NO) & $261^{\circ}$ & $0^{\circ}$ - $360^{\circ}$ \\
  \hline
  $\delta$ (IO) &$277^{\circ}$& $145^{\circ}$ - $391^{\circ}$ \\
\hline
\end{tabular}
\caption{\label{tab1} Current neutrino oscillation parameters from global fits at  3$\sigma$ confidence level \cite{5}. NO(IO) refers to normal (inverted) neutrino mass ordering.}
\end{center}
\end{small}
\end{table}
\section{Numerical analysis}
The experimental constraints on neutrino parameters at 3$\sigma$ confidence levels (CL) are given in
Table \ref{tab1}.
The classification of sixty phenomenologically possible cases of 1TEE and 1TEC is done in the the nomenclature, given by W. Wang in Ref.\cite{17}. All the sixty cases are divided into six categories A, B, C, D and E [Table\ref{tab2}]. In \cite{17}, it is found that the phenomenological results of cases belonging to  1TEC (or 1TEE)  are almost similar to each other due to permutation symmetry.
For the purpose of calculation, we have used the latest experimental data on neutrino mixing angles ($\theta_{12}, \theta_{23},\theta_{13},\delta m^2)$ and mass squared differences ($ \Delta m^2, \delta$) at 3$\sigma$ CL \cite{5}. \\
\begin{figure}[h!]
\begin{center}
\subfigure[]{\includegraphics[width=0.35\columnwidth]{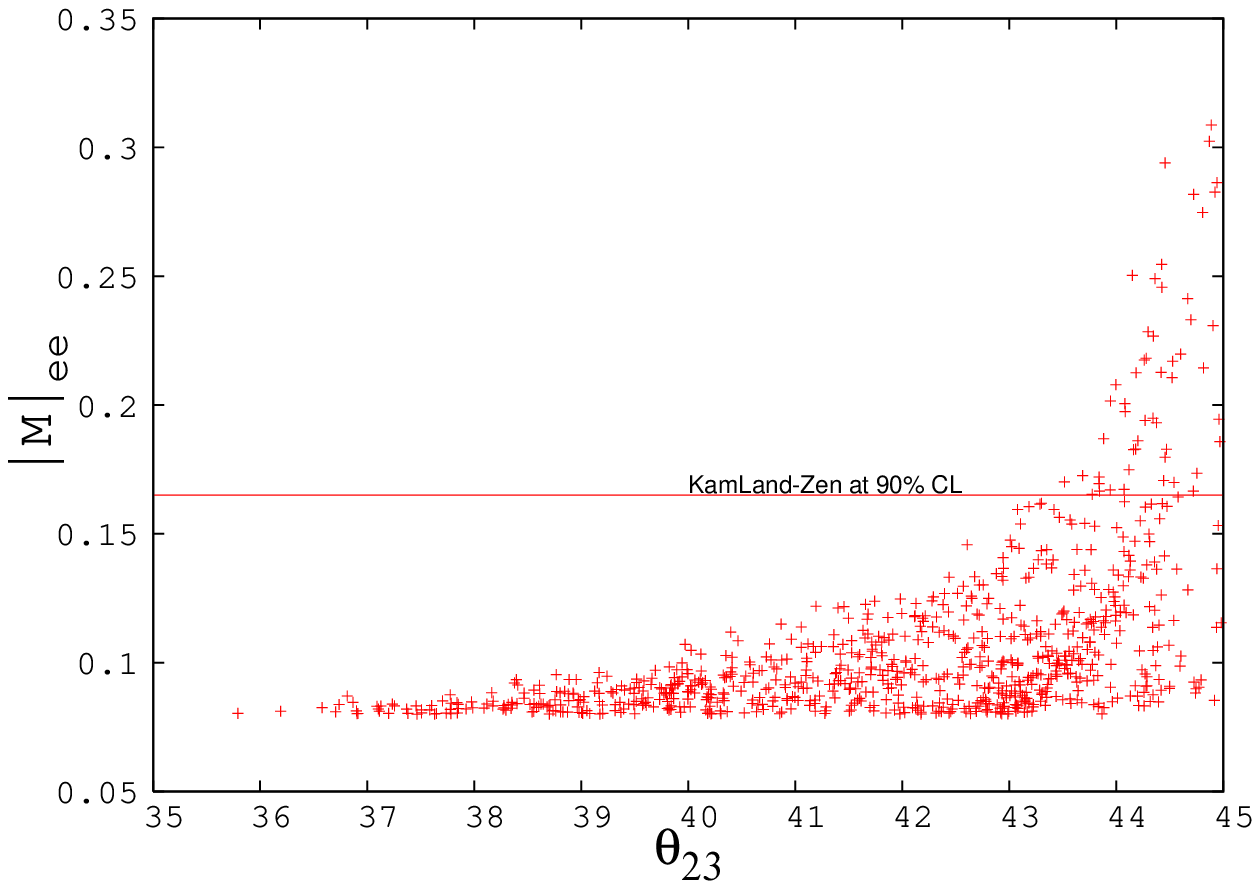}}
\subfigure[]{\includegraphics[width=0.35\columnwidth]{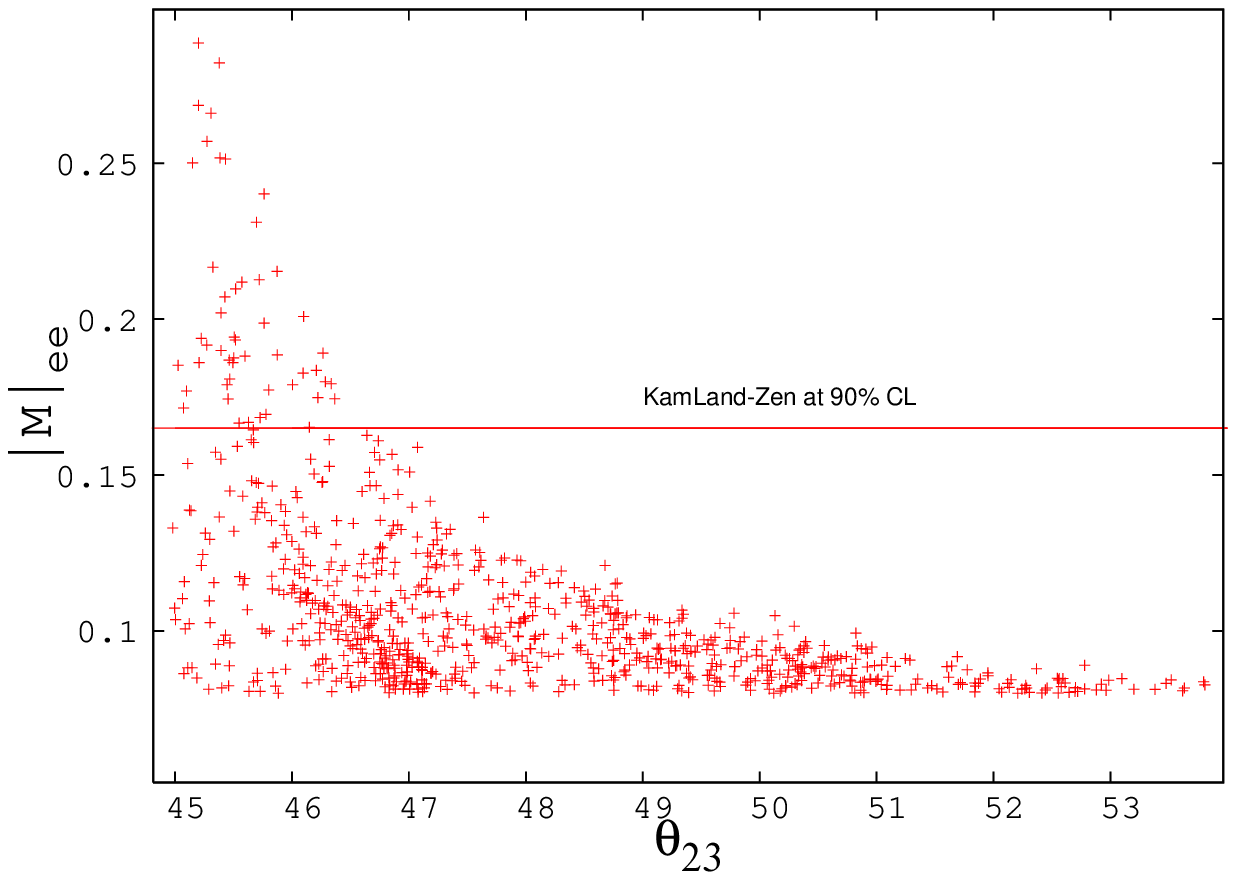}}\\
\subfigure[]{\includegraphics[width=0.35\columnwidth]{HT-C2mee.eps}}
\subfigure[]{\includegraphics[width=0.35\columnwidth]{HT-B2mee.eps}}\\
\caption{\label{fig1}Correlation plots for textures $B_{2}$ [(a) NO (b)IO] and $C_{7}$ for [(c) NO (d) IO] at 3$\sigma$ CL for 1TEE. The symbols have their usual meaning. The horizontal line indicates the upper limit on effective neutrino mass term $|M|_{ee}$ (i.e $|M|_{ee}<0.165eV$) at 90 $\%$ CL, given in KamLAND-Zen experiment \cite{6}.  }
\end{center}
\end{figure}
\begin{figure}[h!]
\begin{center}
\subfigure[]{\includegraphics[width=0.35\columnwidth]{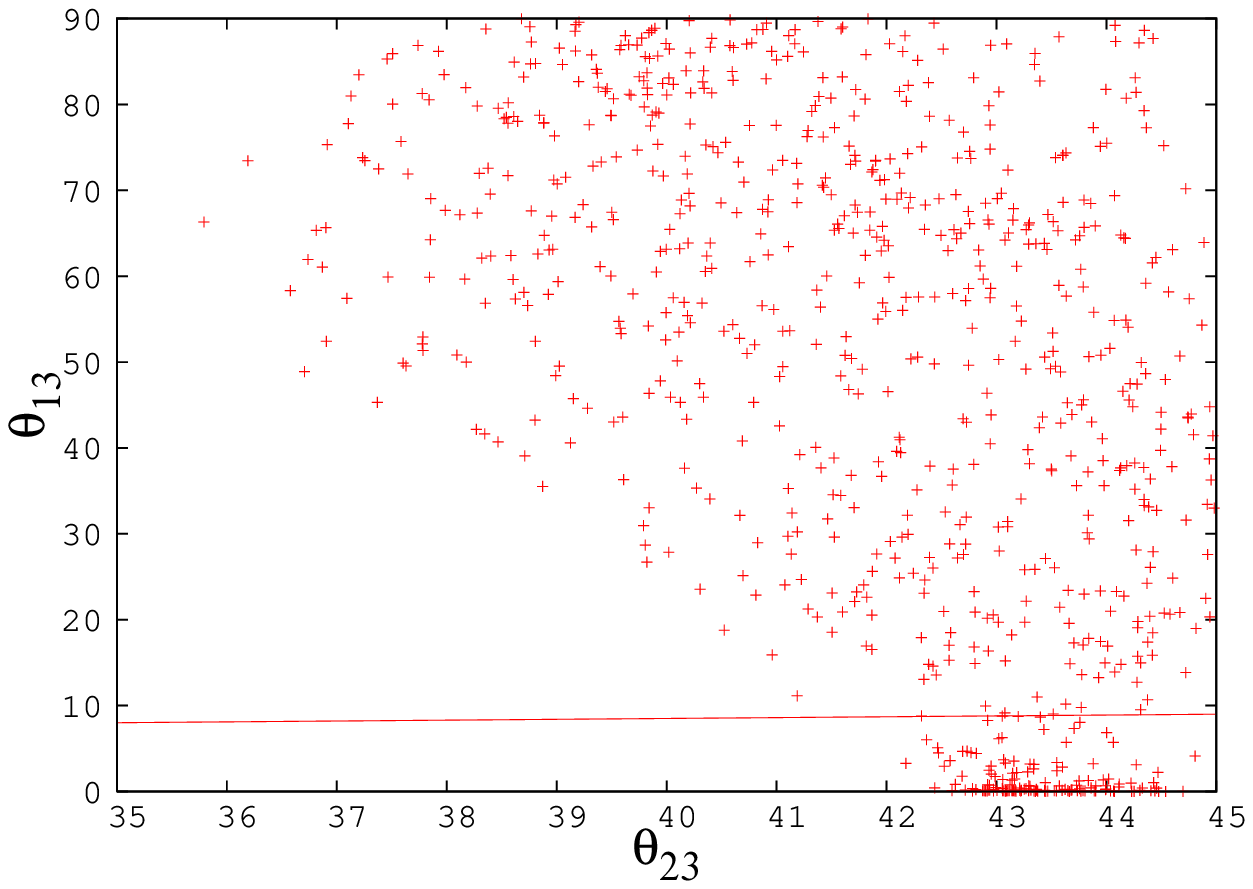}}
\subfigure[]{\includegraphics[width=0.35\columnwidth]{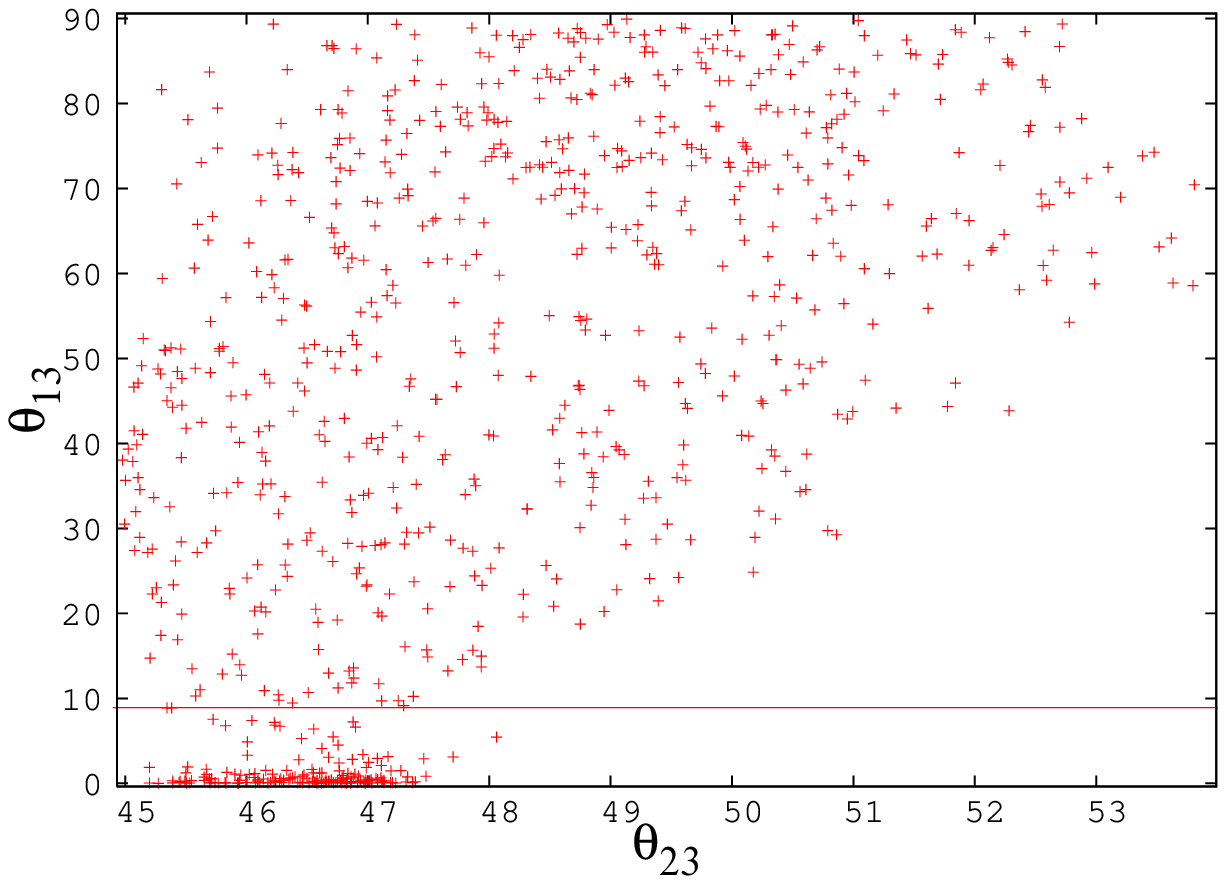}}\\
\subfigure[]{\includegraphics[width=0.35\columnwidth]{HT-C2t23.eps}}
\subfigure[]{\includegraphics[width=0.35\columnwidth]{HT-B2t23.eps}}\\
\caption{\label{fig2}Correlation plots for textures $B_{2}$ [(a) NO (b)IO] and $C_{7}$ for [(c) NO (d) IO] at 3$\sigma$ CL for 1TEE. The symbols have their usual meaning. The horizontal line indicates the upper limit on reactor mixing angle $\theta_{13} <8.9^{0}$, as given in Table \ref{tab1}.}
\end{center}
\end{figure}

\begin{figure}[h!]
\begin{center}
\subfigure[]{\includegraphics[width=0.35\columnwidth]{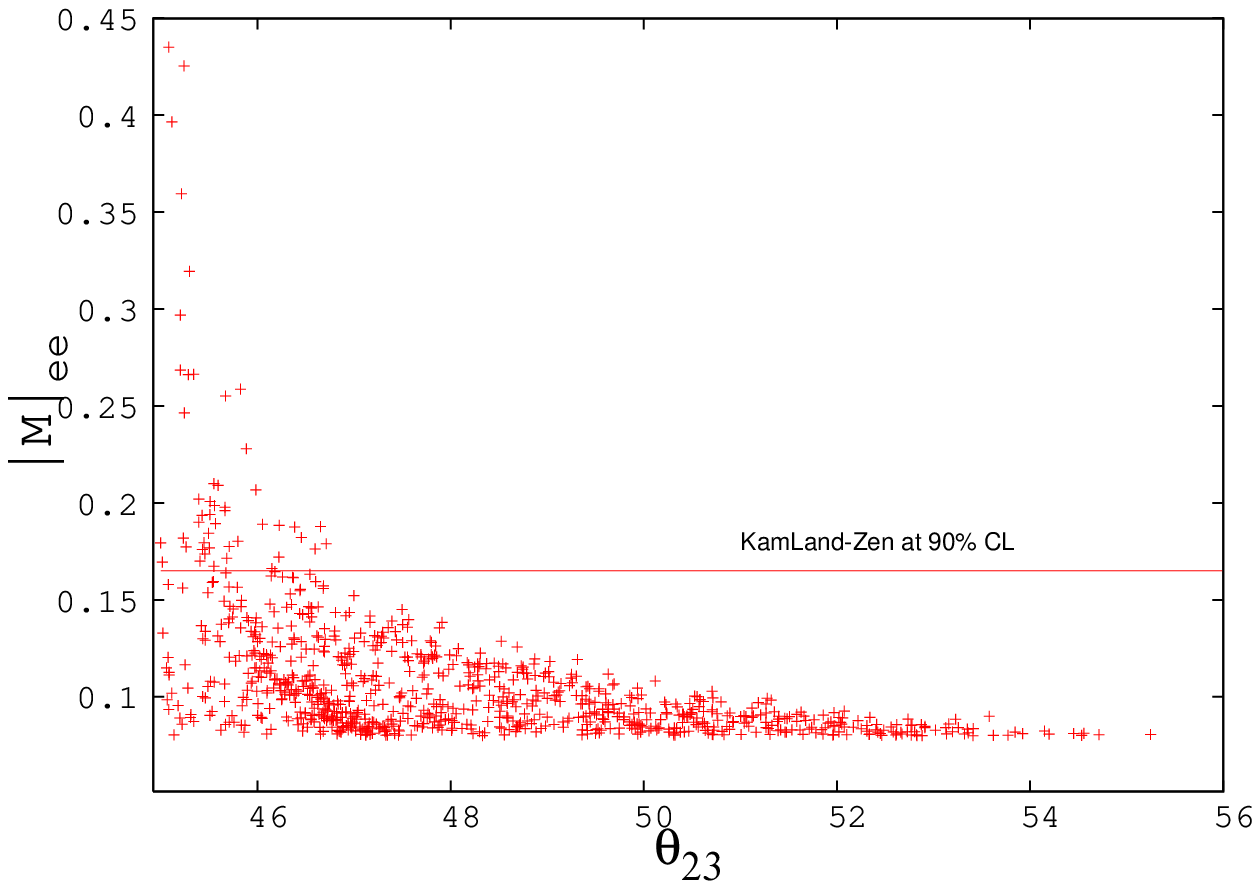}}
\subfigure[]{\includegraphics[width=0.35\columnwidth]{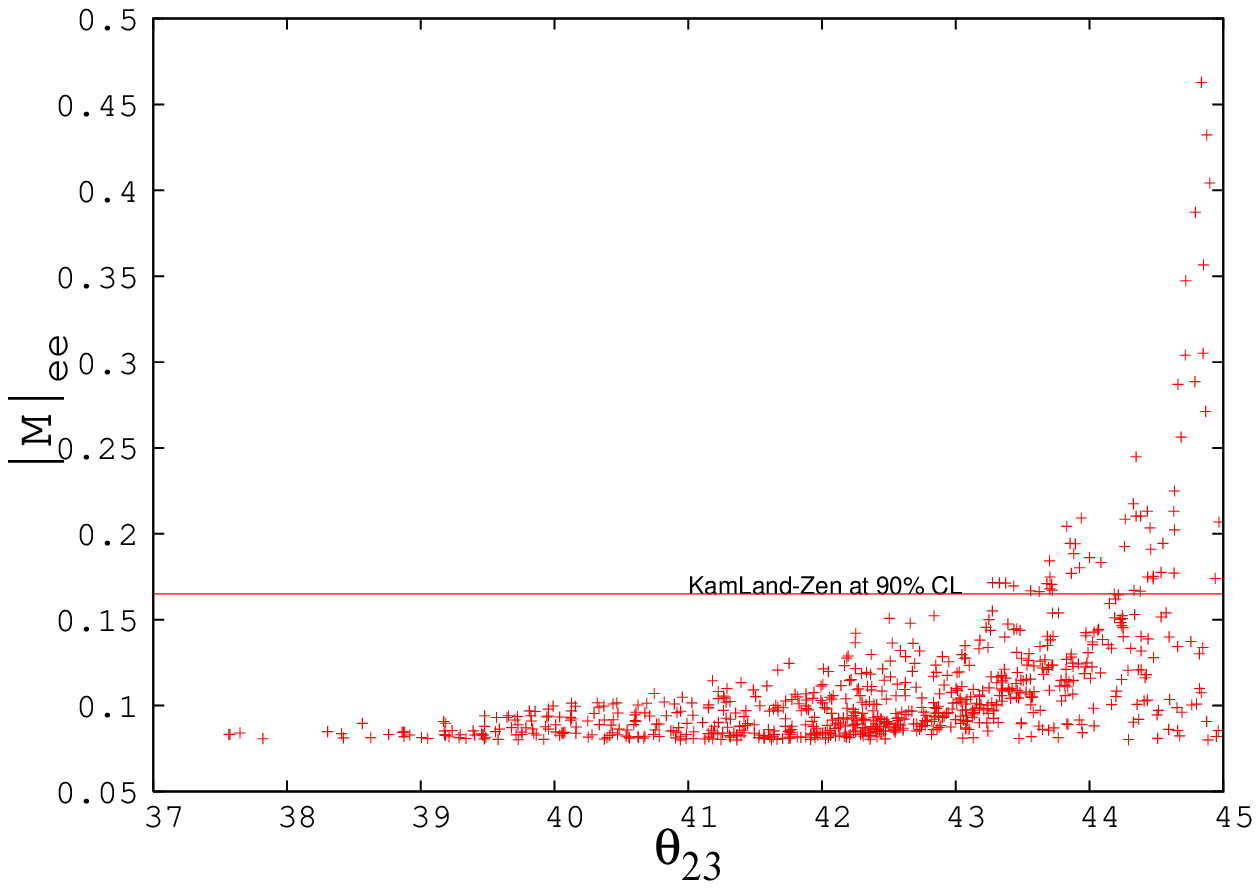}}\\
\subfigure[]{\includegraphics[width=0.35\columnwidth]{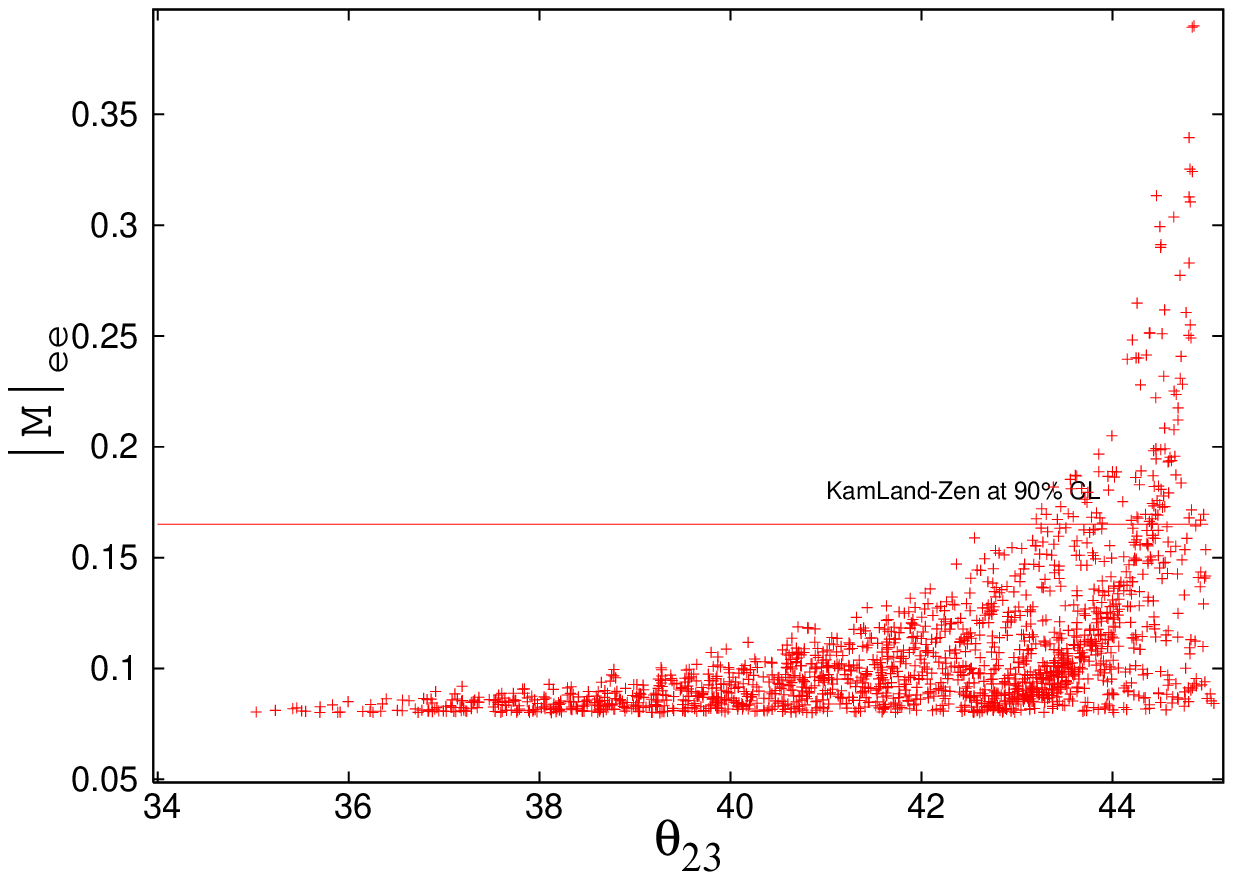}}
\subfigure[]{\includegraphics[width=0.35\columnwidth]{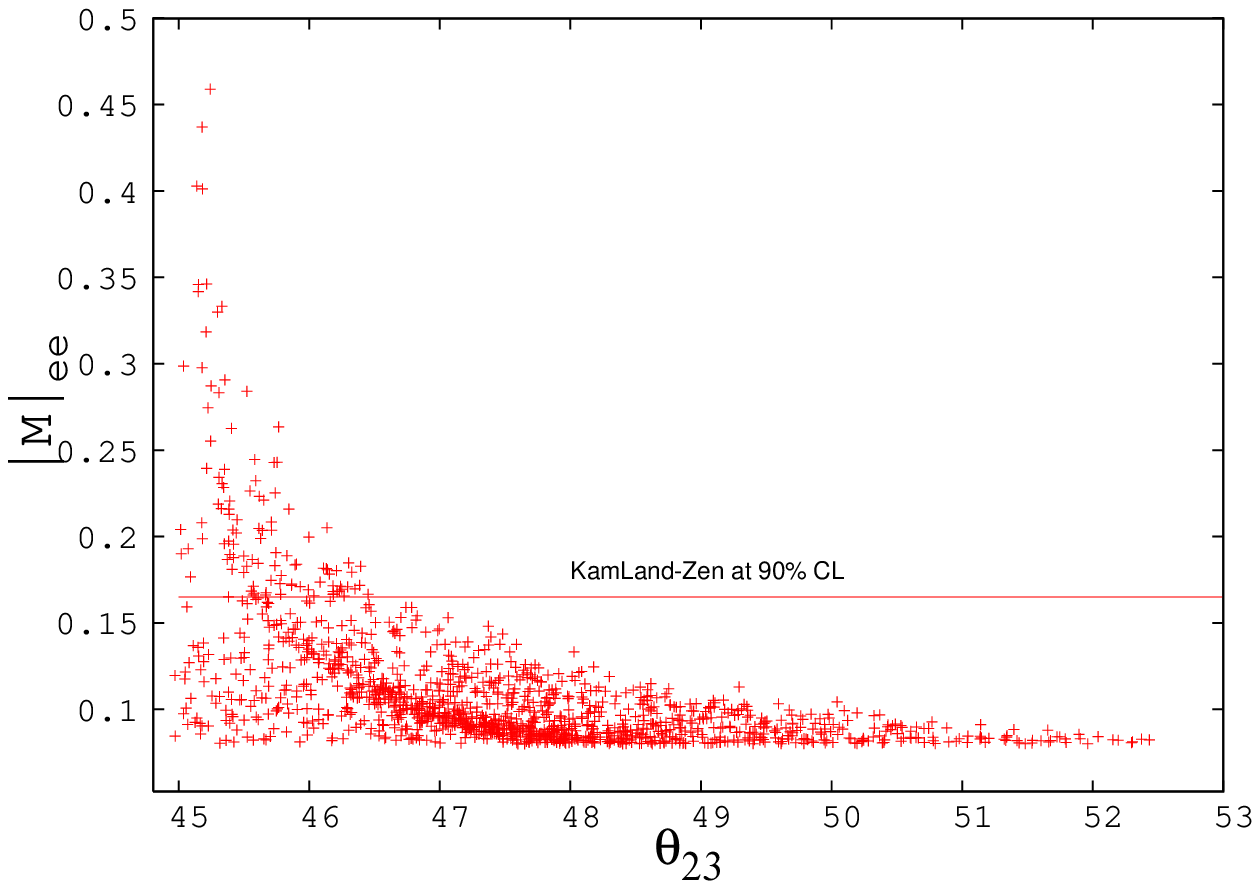}}\\

\caption{\label{fig3}Correlation plots for textures $B_{2}$ [(a) NO (b)IO] and $C_{7}$ for [(c) NO (d) IO] at 3$\sigma$ CL for 1TEC. The symbols have their usual meaning.  The horizontal line indicates the upper limit on effective neutrino mass term $|M|_{ee}$ (i.e $|M|_{ee}<0.165eV$) at 90 $\%$ ($<2\sigma$)CL, given in KamLAND-Zen experiment \cite{6}.}
\end{center}
\end{figure}

\begin{figure}[h!]
\begin{center}
\subfigure[]{\includegraphics[width=0.35\columnwidth]{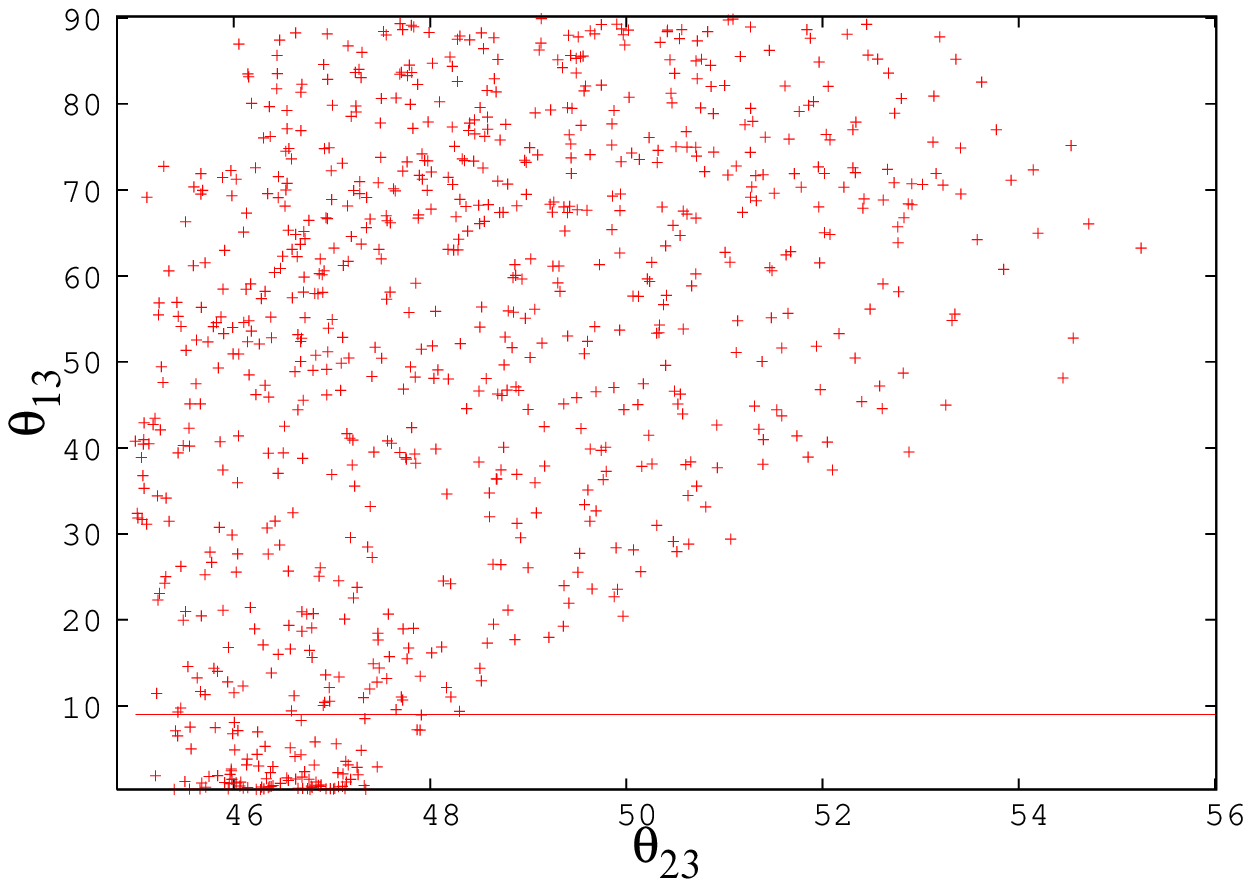}}
\subfigure[]{\includegraphics[width=0.35\columnwidth]{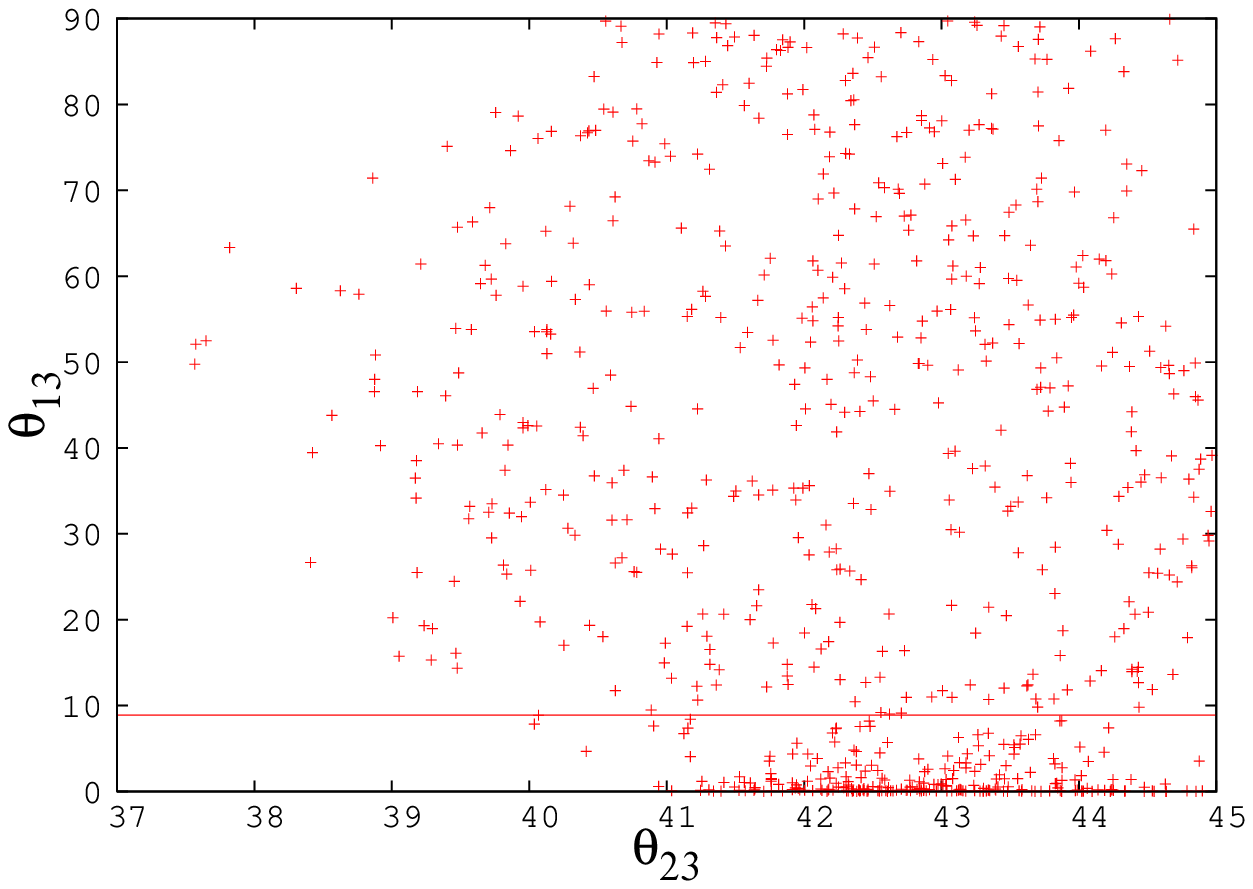}}\\
\subfigure[]{\includegraphics[width=0.35\columnwidth]{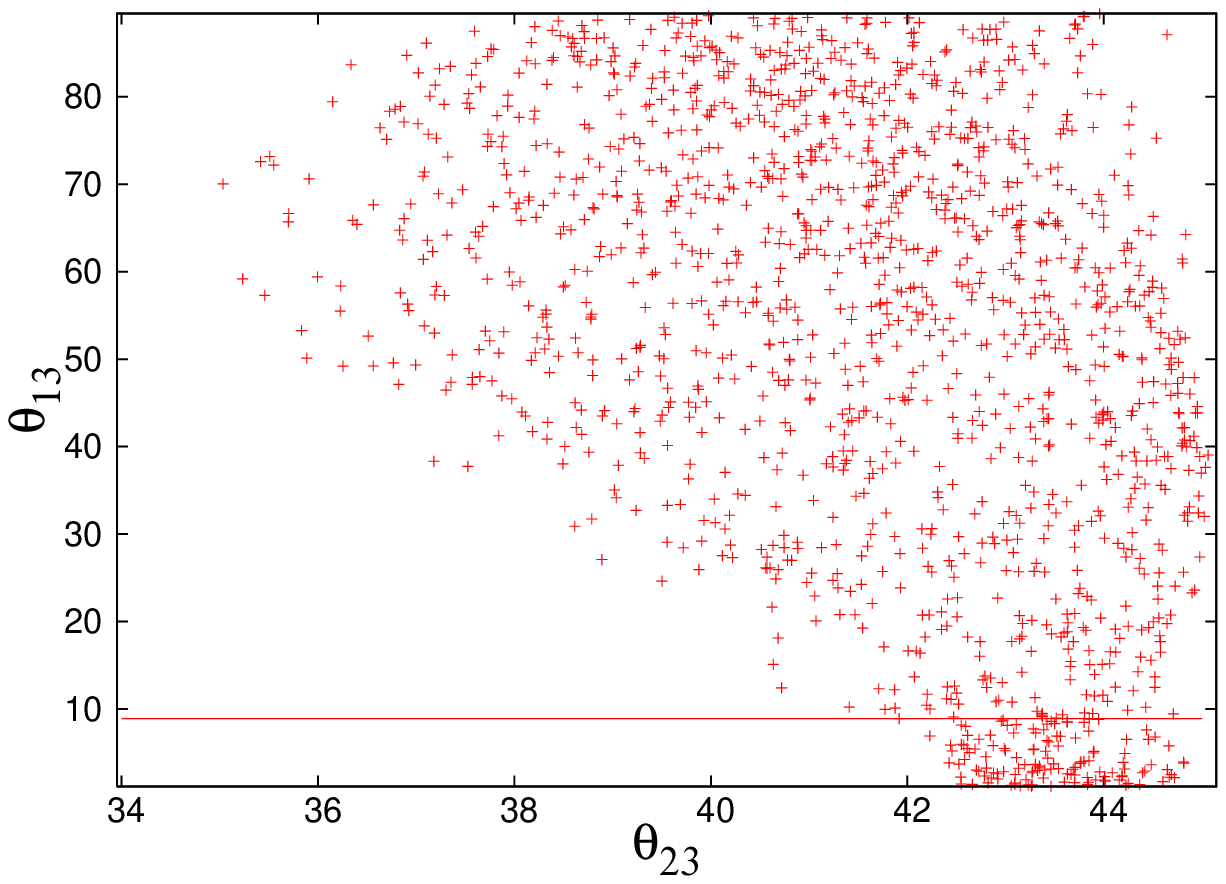}}
\subfigure[]{\includegraphics[width=0.35\columnwidth]{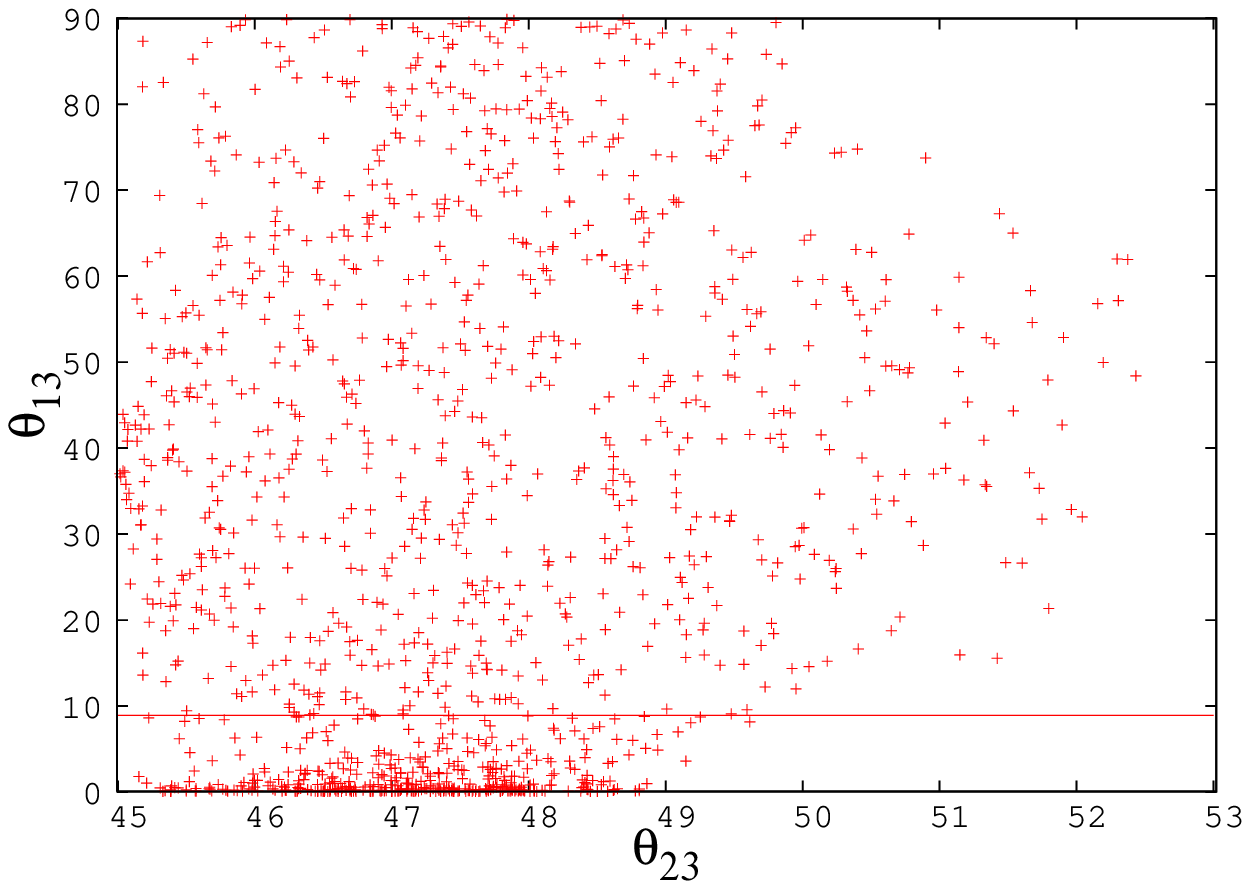}}\\
\caption{\label{fig4}Correlation plots for textures $B_{2}$ [(a) NO (b)IO] and $C_{7}$ for [(c) NO (d) IO] at 3$\sigma$ CL for 1TEC. The symbols have their usual meaning.  The horizontal line indicates the upper limit on reactor mixing angle $\theta_{13} <8.9^{0}$, as given in Table \ref{tab1}.}
\end{center}
\end{figure}
\subsection{Near maximal atmospheric mixing for 1TEE and 1TEC texture structures}
\begin{figure}[h!]
\begin{center}
\subfigure[]{\includegraphics[width=0.35\columnwidth]{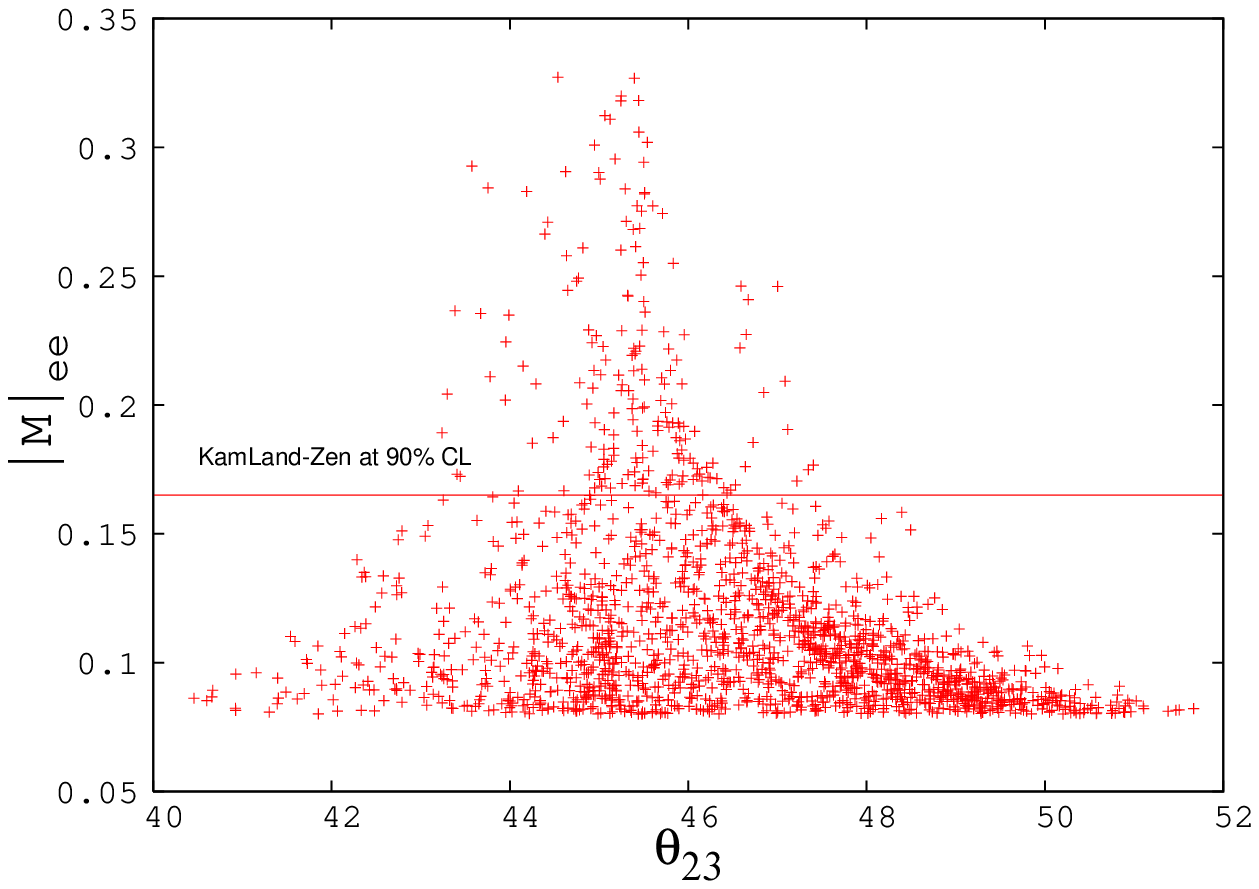}}
\subfigure[]{\includegraphics[width=0.35\columnwidth]{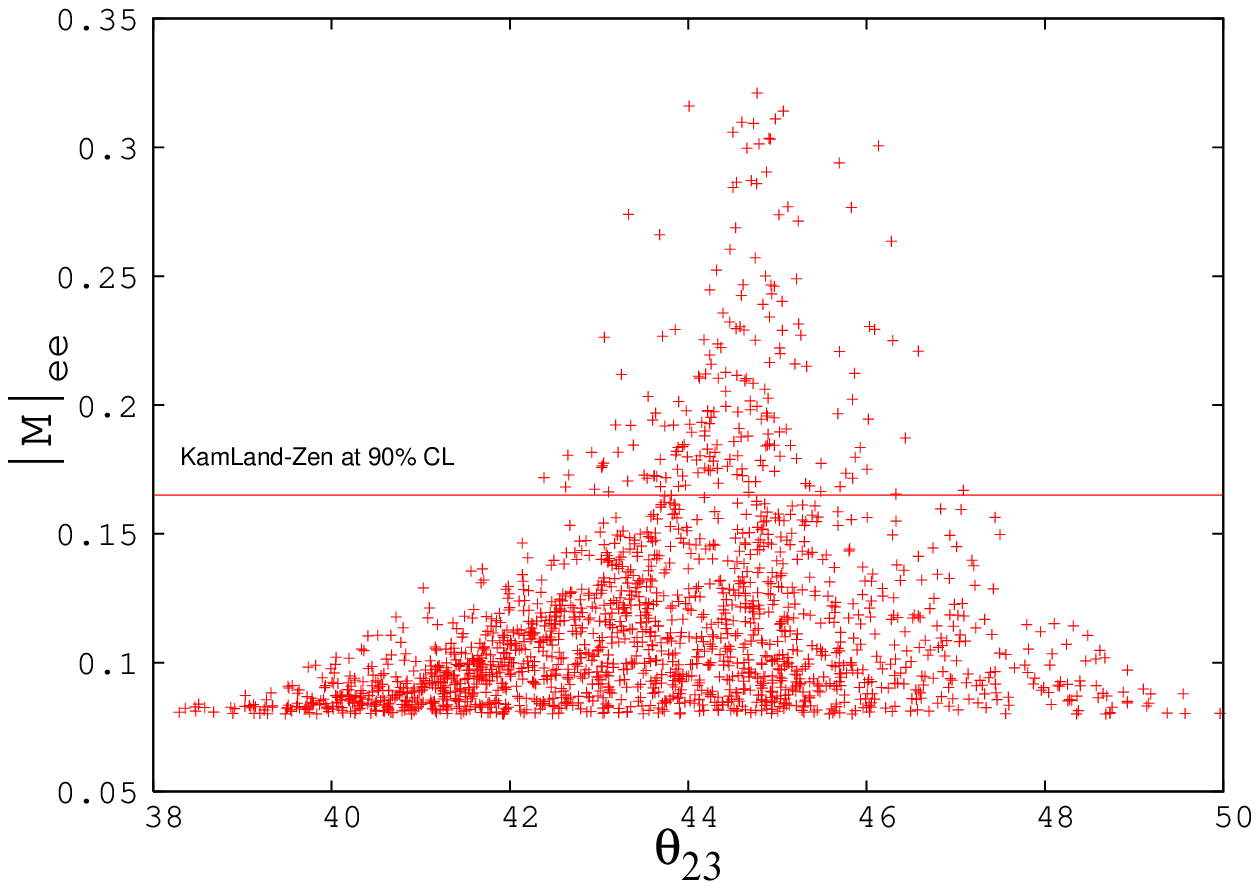}}\\
\subfigure[]{\includegraphics[width=0.35\columnwidth]{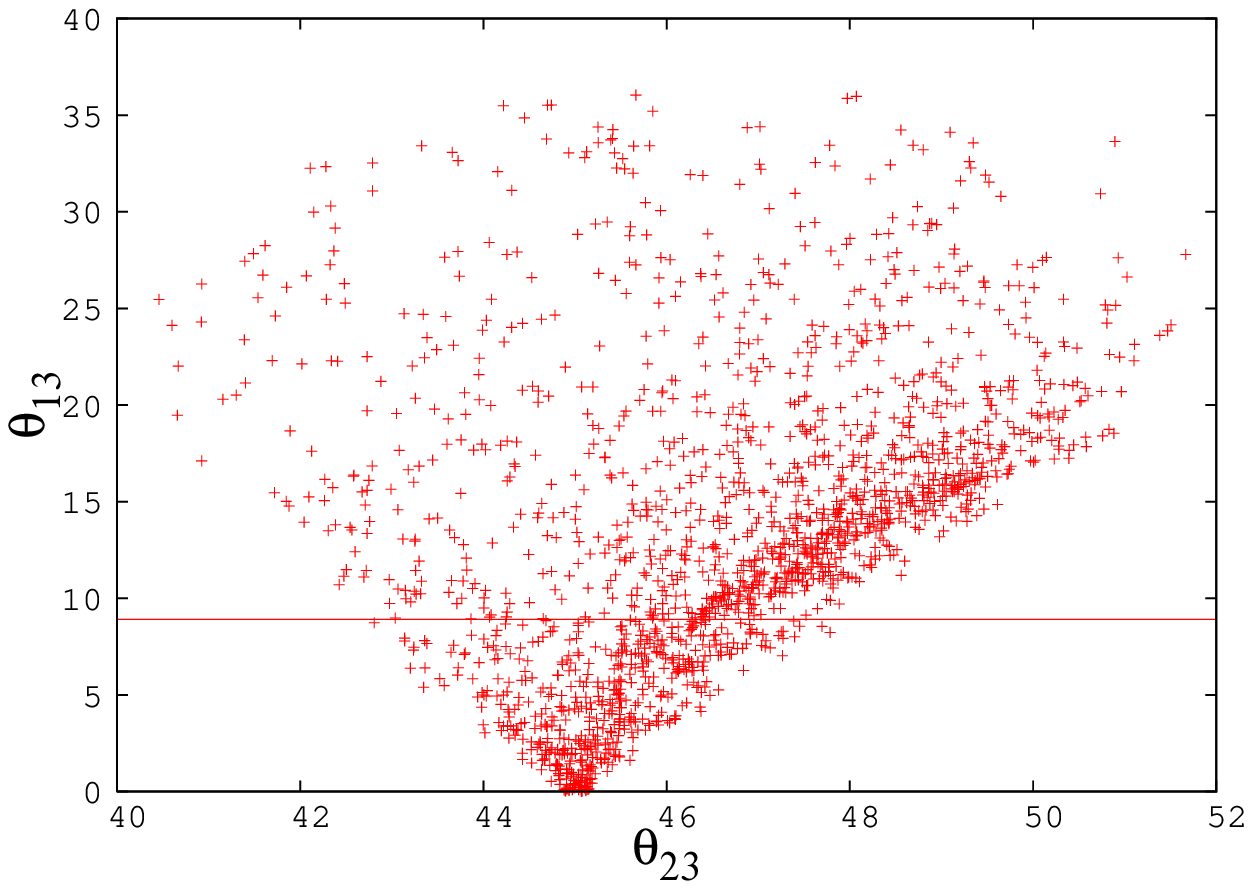}}
\subfigure[]{\includegraphics[width=0.35\columnwidth]{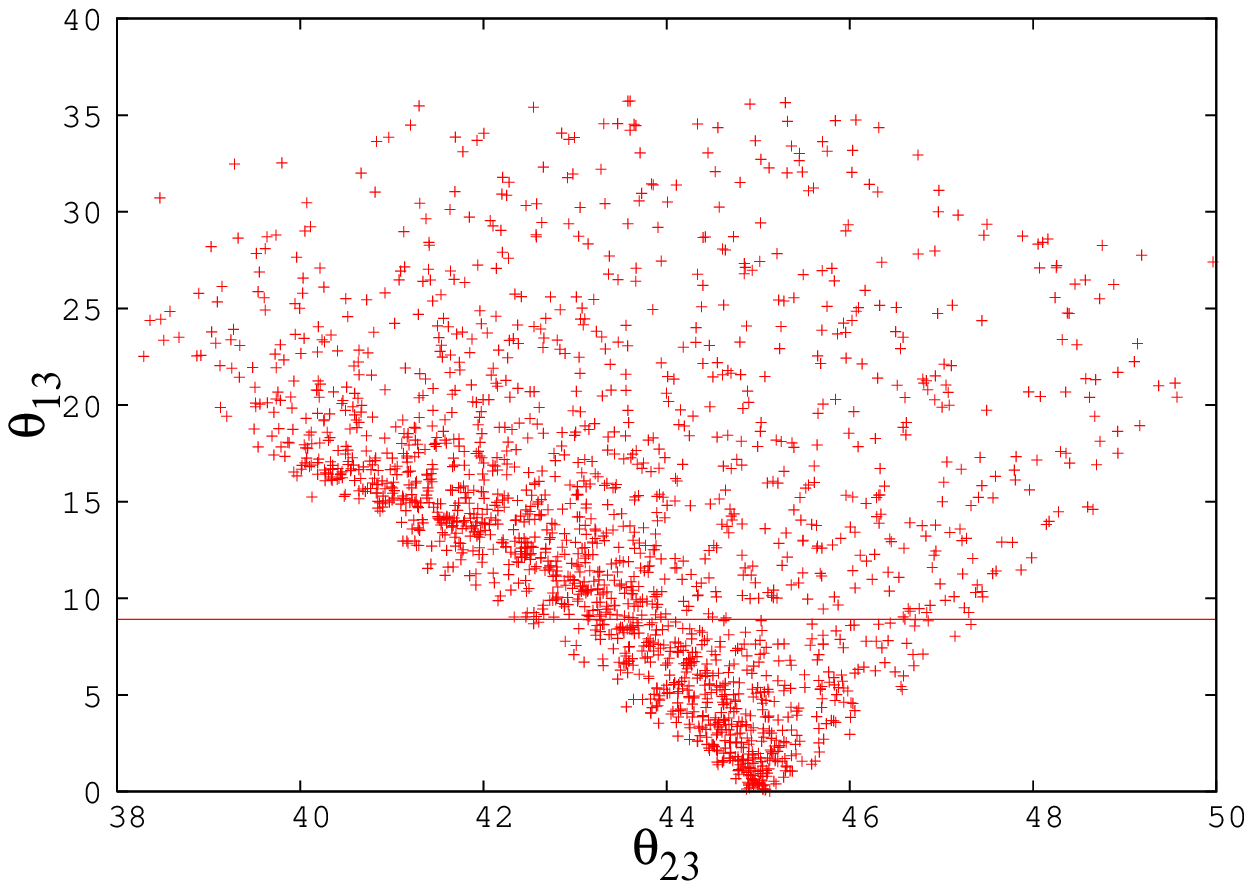}}\\
\caption{\label{fig5}Correlation plots for textures $D_{3}$ [(a),(c)] and  $F_{4}$ [(b),(d)] with IO for 1TEE at 3$\sigma$ CL. The symbols have their usual meaning.  In Figs. (a) and (b), colored horizontal line indicates the upper limit on effective neutrino mass term $|M|_{ee}$ (i.e $|M|_{ee}<0.165eV$) at 90 ($<2\sigma$)$\%$ CL, given in KamLAND-Zen experiment \cite{6}.  In Figs.(c) and (d), we have shown the  upper limit on reactor mixing angle $\theta_{13}$. }
\end{center}
\end{figure}

\begin{figure}[h!]
\begin{center}
\subfigure[]{\includegraphics[width=0.35\columnwidth]{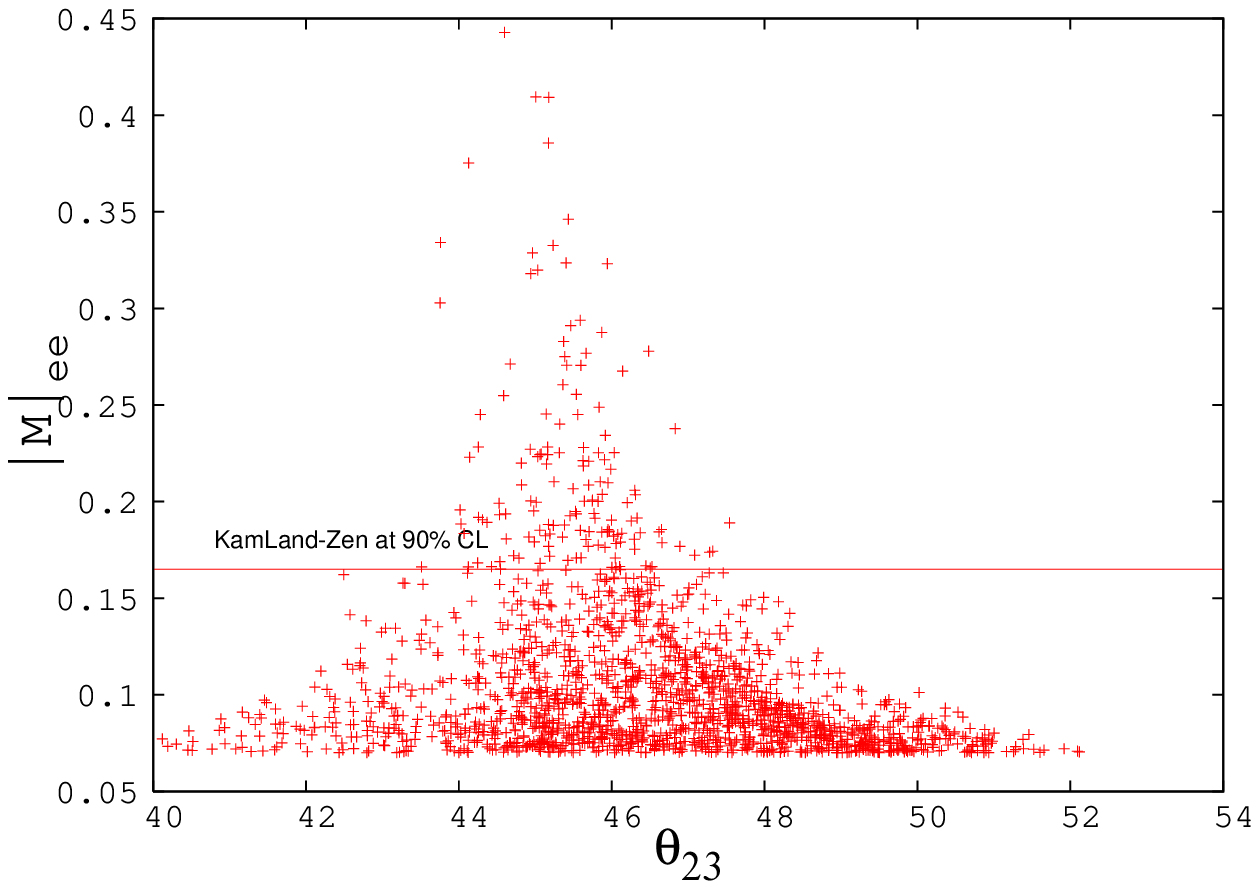}}
\subfigure[]{\includegraphics[width=0.35\columnwidth]{iHT-F4mee.eps}}\\
\subfigure[]{\includegraphics[width=0.35\columnwidth]{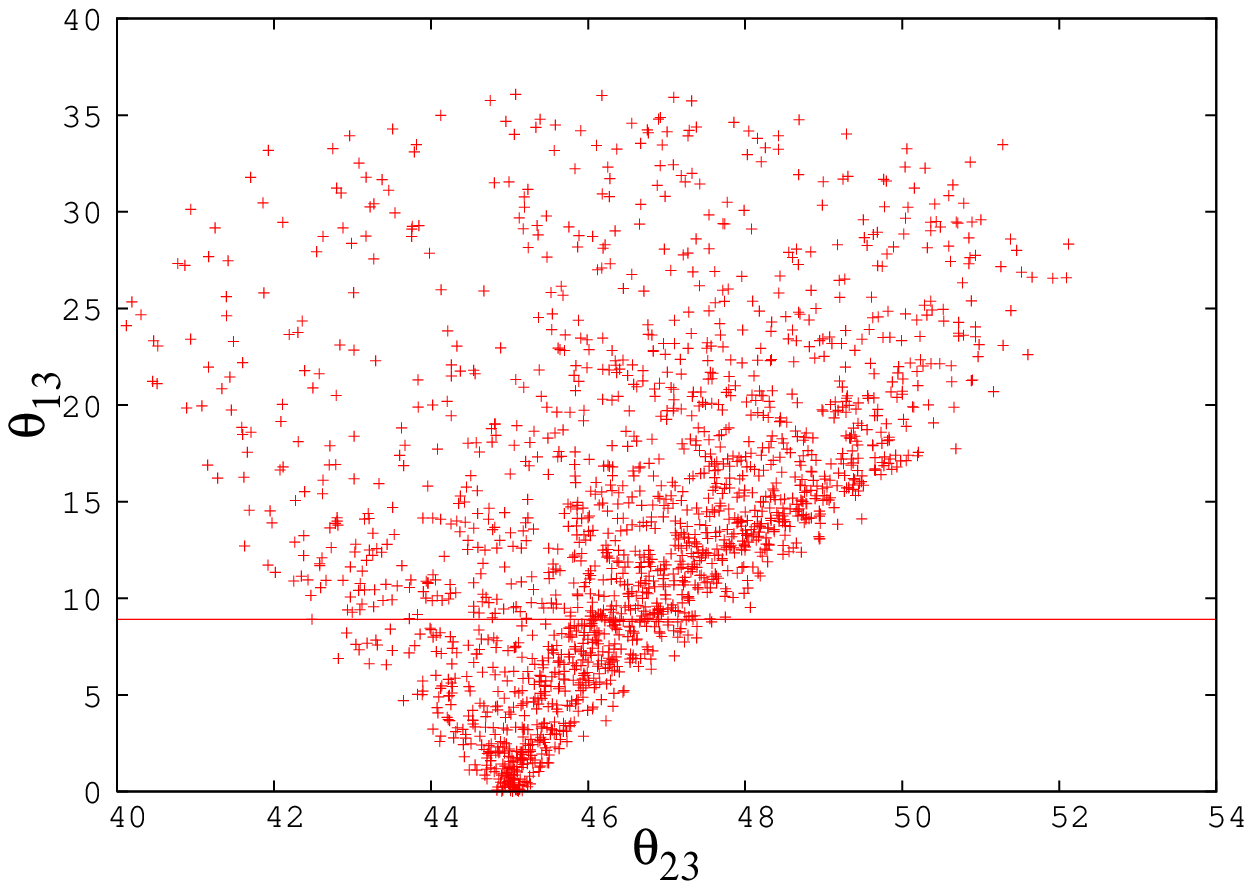}}
\subfigure[]{\includegraphics[width=0.35\columnwidth]{iHT-F4t23.eps}}\\
\caption{\label{fig6}Correlation plots for textures $D_{3}$ [(a),(c)] and  $F_{4}$ [(b),(d)] with NO for 1TEC at 3$\sigma$ CL. The symbols have their usual meaning.  In Figs. (a) and (b), colored horizontal line indicates the upper limit on effective neutrino mass term $|M|_{ee}$ (i.e $|M|_{ee}<0.165eV$) at 90 ($<2\sigma$)$\%$ CL, given in KamLAND experiment \cite{6}.  In Figs. (c) and (d), we have shown the  upper limit on reactor mixing angle $\theta_{13}$. }
\end{center}
\end{figure}

As a first step of the analysis, all the sixty cases of 1TEE and 1TEC have been investigated in the limit of large $|M|_{ee}$.  For the analysis, we have incorporated the assumptions of Refs. \cite{18, 19}, wherein authors have considered the lower bound on $|M|_{ee}$  to be large (i.e. $|M|_{ee} > 0.08 eV$).  The upper bound on $|M|_{ee}$ is choosen to be  more conservative  i.e. $|M|_{ee} < 0.5eV$  at 3$\sigma$ CL \cite{9}. The input parameters ($\theta_{12}, \theta_{23},\theta_{13},\delta m^2, \Delta m^2, \delta$) are generated by the method of random number generation. The three neutrino mixing angles and Dirac-type CP-violating phase $\delta$ are varied between $0^{0}$ to $90^{0}$ and $0^{0}$ to $360^{0}$, respectively.  However, the mass-squared differences ($\delta m^2, \Delta m^2$) are varied randomly within their 3$\sigma$ experimental range \cite{5}.  For the numerical analysis, we follow the same procedure as discussed in \cite{13}.
The main results and discussion are summarized as follows: 
  
In Fig. \ref{fig1}, Fig. \ref{fig2}, Fig. \ref{fig3}, Fig. \ref{fig4}, Fig. \ref{fig5}, Fig. \ref{fig6}, it is explicitly shown that the octant of $\theta_{23}$ is well restricted for $B_{2}, C_{7}, D_{3}, F_{4}$ of 1TEE and 1TEC texture structures, respectively. However, for the remaining cases, the value of $\theta_{23}$ is unconstrained like other oscillation parameters. Apart from restricting the octant of $\theta_{23}$, the analysis also ensures the quasi degenerate mass ordering for these cases similar to the observation of Refs.\cite{18,19,20}. From Fig. [\ref{fig1}(a, b)] and Fig. [\ref{fig3}(a, b)], it is clear that for increasing value of $|M|_{ee}$, atmospheric mixing angle $\theta_{23}$ approaches to maximal value for the structure $B_{2}$ of 1TEE and 1TEC for both normal (NO) as well as inverted (IO) ordering. In Fig. \ref{fig2} and Fig. \ref{fig4}, it is explicitly shown that for cases $B_{2}$ and $C_{7}$ the quadrant of $\theta_{23}$ is already decided without the experimental input of the mixing angles.
For 1TEE we have  $\theta_{23}<45^{0}$ for NO, while $\theta_{23}>45^{0}$ for IO, whereas for 1TEC, $\theta_{23}>45^{0}$ for NO, while $\theta_{23}<45^{0}$ for IO [Fig. \ref{fig2}(a, b), Fig. \ref{fig4}(a, b)]. Clearly  the correlation plots of case $B_{2}$  are  indistinguishable for 1TEE and 1TEC, if neutrino mass ordering is not considered as also pointed out earlier. Similar conclusion can be drawn for structure $C_{7}$ since both are related through 2-3 exchange symmetry [Fig. \ref{fig2}(c, d), Fig. \ref{fig4}(c, d)]. Apart from the prediction of near maximality of $\theta_{23}$,  cases $B_{2}$ and $C_{7}$ also predict $\delta \simeq 90^{0}, 270^{0}$ for 1TEE and 1TEC respectively, if experimental range of mixing angles are considered Table [\ref{tab2}]. Figs. \ref{fig2} [(a), (c)] for NO and Figs. \ref{fig2}(b), (d)] for IO, depict the 2-3 interchange symmetry between the cases $B_{2}$ and $C_{7}$ for 1TEE. Similar phenomenological observation is shown for 1TEC in Fig. \ref{fig4}(a, c) and Fig. \ref{fig4} (b, d), respectively.

Similarly, cases $D_{3}$ and $F_{4}$ of 1TEE  also predict near maximal atmospheric mixing angle ($\theta_{23}$) for IO [Fig. \ref{fig5}(a, b)]. Interestingly the parameter space of reactor mixing angle $\theta_{13}$  is found to be constrained between $0^{0}$ and $35^{0}$ [Figs. \ref{fig5}(c), (d)]. In Figs. \ref{fig5}(c, d), it is clear that  for the allowed experimental range of $\theta_{13}$ ($8.5^{0}-9.8^{0}$), $\theta_{23}$ inches closer to $45^{0}$. Similar predictions have been noted for cases $D_{3}$ and $F_{4}$ of 1TEC, however for normal mass ordering (NO) [Figs. \ref{fig6}(a, b, c, d)].  
\subsection{Comparing the results for 1TEE and 1TEC texture structures}
In this subsection, we compare the results of all the viable structures of 1TEE and 1TEC in neutrino mass matrix. It is worthwhile to mention that the present refinements of the experimental data does not limit the number of viable cases in 1TEE and 1TEC  textures repectively. The number of viable cases obtained are same as predicted in Refs.\cite{14} and \cite{17} for 1TEE and 1TEC, respectively.
For executing the analysis, we vary the allowed ranges of three neutrino mixing angles ($\theta_{12},\theta_{23},\theta_{13}$) and mass squared differences ($\delta m^2, \Delta m^2$) within their 3$\sigma$ confidence level.  To facilitate the comparison, we have encapsulated the the predictions regarding three CP violating phases ($\rho, \sigma, \delta$) and neutrino masses $m_{1,2,3}$ for all the allowed texture structures of 1TEE and 1TEC respectively [Table \ref{tab3}, \ref{tab4}, \ref{tab5}, \ref{tab6}].   

\textbf{Category A}: In Category A, there are 10 possible cases out of which only four ($A_{1,4,5,6}$) are allowed for 1TEE at 3$\sigma$ CL, and in addition, inverted mass ordering (IO) is ruled out for all these cases. On the other hand, only three ($A_{1,4,6}$) are allowed for 1TEC  with current oscillation data, while normal mass ordering (NO) is ruled out for these cases. For 1TEE,  $\rho, \sigma, \delta$ remain unconstrained, however for 1TEC, only $\delta$ remains unconstrained, while Majorana phases ($\rho, \sigma)$ are restricted near $0^{0}$ pertaining to viable cases. From Table \ref{tab3}, it is clear that lower bound on lowest neutrino mass ($m_{1}$ (NO) or $m_{3}$ (IO)) is nearly equal or less than 1meV for 1TEE and 1TEC.
 
\textbf{Category B (C)}: In Category B, all the ten possible cases are allowed for both 1TEE and 1TEC, respectively at 3$\sigma$ CL, however cases $B_{1,6,7}$ allow only NO for 1TEE, while the same  allow only IO for 1TEC [Table \ref{tab4}]. Cases $B_{2,3,4,5,8,9,10}$ allow both NO as well as IO for 1TEE and 1TEC, respectively. As mentioned in Ref. \cite{17}, cases of Category B are related to the cases belonging to Category C through permutation symmetry, therefore we can obtain the results for Category C from B. We find that cases $C_{1,2,3}$ allow only NO for 1TEE, while the same allow IO for 1TEC.  \\
Textures $B_{1}$(NO),$ B_{3}$ (NO,IO), $B_{5}$(NO, IO), $B_{6}$(NO), $B_{7}$(NO), $B_{8}$(NO), $B_{10}$ (NO), $C_{1}$(NO), $C_{2}$(NO), $C_{3}$(NO), $C_{4}$(NO,IO), $C_{6}$(NO,IO), $C_{8}$(NO), $C_{10}$(NO) held nearly no constraint on Dirac CP violating phase ($\delta$) for 1TEE and 1TEC respectively, however with opposite neutrino mass ordering [Table \ref{tab4}].  Only cases $B_{2}$ (NO), $B_{4}$ (IO), $C_{7}$(NO), $C_{5} $(IO) for 1TEE, and $B_{2}$ (IO),$B_{4}$(NO), $C_{7}$(IO), $C_{5}$ (NO) for 1TEC show significant reduction in the parameter space of $\delta$.  It is found that $\delta$ is restricted near $ 90^{0}$ and $ 270^{0}$  for 1TEE and 1TEC, respectively  [Table \ref{tab4}]. These predictions are significant considering the latest hint on $\delta$ near $270^{0}$ \cite{5}.
Therefore all the above cases discussed are almost indistinguishable for 1TEE and 1TEC, if neutrino mass ordering is not considered.

 \textbf{Category D (F)}:  All the ten possible cases belonging to Category D are acceptable with neutrino oscillation data at 3$\sigma$ CL for 1TEE and 1TEC, respectively [Table \ref{tab5}]. However cases $D_{1,2,4,5,6,7,9}$  favor both NO and IO for 1TEE and 1TEC, while $D_{3}, D_{8}$ and $D_{10}$ are acceptable only for IO in case of 1TEE, however same cases are allowed for NO in case of 1TEC. Similarly, the results for cases belonging to Category F can be derived from Category D.\\
 Cases $D_{1}$(IO), $D_{2}$ (IO), $D_{3}$(IO), $D_{4}$(IO), $D_{5}$(NO, IO), $D_{6}$(IO), $D_{7}$(NO, IO), $D_{8}$(IO), $D_{9}$(NO, IO), $D_{10}$(IO), $F_{1}$(IO),$F_{2}$ (IO), $F_{3}$(IO), $F_{4}$(IO), $F_{5}$(NO, IO), $F_{6}$(IO), $F_{7}$(NO, IO), $F_{8}$(IO), $F_{9}$(NO, IO), $F_{10}$(IO) predict literally no constraints on $ \delta$ for 1TEE. These cases give identical predictions for 1TEC, however for opposite mass ordering. On the other hand, for $D_{1}$(NO), $D_{2}$(NO), $D_{4}$(NO), $D_{6}$(NO), $F_{1}$(NO),$ F_{2}$(NO), $F_{4}$(NO), $F_{6}$(NO) $\delta$ is notably constrained for 1TEE, and same is true for 1TEC, however for opposite mass ordering.
 
\textbf{Category E}: In Category E, all the ten possible cases are allowed for 1TEE at 3$\sigma$ CL, while only nine other than $E_{5}$ are acceptable in case of 1TEC [Table \ref{tab6}]. Cases $E_{1,2,3,4,6,7,8,9}$ allow only inverted mass ordering (IO) for 1TEE, while same textures allow only normal mass ordering (NO) for 1TEC. Cases $E_{5}$ and $E_{10}$ allow both NO and IO for 1TEE, however $E_{5}$ is ruled out for both NO and IO for 1TEC at 3$\sigma$ CL. Similar to cases belonging to Category D,  $E_{1,2,3,4,5,6,7,8,9}$ (IO) cover full range of $\rho, \sigma, \delta$  for 1TEE,  whereas same cases (except $E_{5}$) give identical predictions for 1TEC, however for NO. For $E_{5}$(NO) and $E_{10}$ (NO), phases $\rho, \sigma, \delta$  are somewhat restricted at 3$\sigma$ CL for 1TEE, while only for $E_{10}$ (IO), the parameter space of $\rho, \sigma, \delta$  seem to be restricted for 1TEC [Table \ref{tab6}]. \\

To summarize our discussion, we have investigated all the viable cases of 1TEE and 1TEC texture structures in the limit of large effective neutrino mass $|M|_{ee}$. It is found that only four cases are able to produce near maximal atmospheric mixing for 1TEE and 1TEC, respectively. However, the predictions remain true irrespective of the experimental data on solar and reactor mixing angle. The observation also hints towards the indistinguishable feature of 1TEE and 1TEC texture stuctures, however for opposite mass ordering. In order to depict the indistinguishablity, we have carried out a comparitive study of 1TEE and 1TEC texture structures using the current experimental data at 3$\sigma$ CL. From our discussion we find that most of the cases belonging to 1TEE and 1TEC are almost indistinguishable as far as the neutrino oscillation parameters are concerned, however with opposite neutrino mass ordering. The indistinguishable nature of 1TEE and 1TEC is more prominent for quasi degenerate mass ordering. For the cases where lower bound on lowest neutrino mass is very small ($<1meV$), there is noticeable deviation in the predictions for 1TEE and 1TEC [Table \ref{tab3}, \ref{tab4}, \ref{tab5} and \ref{tab6}]. This point is also discussed by Liao et. al. in Ref. \cite{21}. In addition, the parameter space of $\delta$ for  most of the cases belonging to 1TEE and 1TEC remain unrestricted, while only eight cases show maximal restriction for $\delta$.  Since no presently feasible experiment has been able to determine the neutrino mass ordering, therefore we cannot distinguish 1TEE and 1TEC structures using the present oscillation data.  
However, the currently running and forthcoming neutrino experiments aimed at distinguishing the mass ordering of neutrinos will test our phenomenological results. Also the ongoing and future neutrinoless double beta decay experiments are capable of measuring $|M|_{ee}$ term, which would, in turn either confirm or rule out our assumption of large $|M|_{ee}$.

\section*{Conflicts of Interest}
The authors declare that there are no conflicts of interest regarding the publication of this paper.

\section*{Acknowledgment}

The author would like to thank the Director, National Institute of Technology Kurukshetra, for providing necessary facilities to work.  \\

\newpage

\begin{table}[ht]
\begin{center}
\begin{footnotesize}
\resizebox{14cm}{!}{
\begin{tabular}{|c|c|c|c|c|c|c|}
\hline & A & B & C & D & E & F  \\
 \hline 1 & $\left(
\begin{array}{ccccccc}
0 & \Delta & \Delta \\
\Delta & \times & \times\\
\Delta& \times & \times \\
\end{array}
\right)$ & $\left(
\begin{array}{ccccccc}
\Delta& 0 & \Delta \\
0 & \times & \times\\
\Delta& \times & \times \\
\end{array}
\right)$ & $\left(
\begin{array}{ccccccc}
    \Delta& \Delta & 0 \\
  \Delta & \times & \times\\
  0& \times & \times \\
\end{array}
\right)$ & $\left(
\begin{array}{ccccccc}
    \Delta& \Delta & \times \\
  \Delta & 0 & \times\\
  \times& \times & \times\\
\end{array}
\right)$& $\left(
\begin{array}{ccccccc}
    \Delta& \Delta & \times \\
  \Delta & \times & 0\\
  \times& 0 & \times \\
\end{array}
\right)$& $\left(
\begin{array}{ccccccc}
    \Delta& \Delta & \times \\
  \Delta & \times & \times\\
  \times& \times & 0 \\
\end{array}
\right)$ \\

\hline 2 &$\left(
\begin{array}{ccccccc}
0 & \Delta & \times \\
\Delta & \Delta & \times\\
\times& \times& \times \\
\end{array}
\right)$ & $\left(
\begin{array}{ccccccc}
\times & 0 & \Delta \\
0 & \Delta & \times\\
\Delta& \times & \times \\
\end{array}
\right)$ & $\left(
\begin{array}{ccccccc}
    \Delta & \times & 0 \\
  \times & \Delta & \times\\
  0& \times & \times \\
\end{array}
\right)$ & $\left(
\begin{array}{ccccccc}
    \Delta& \times & \Delta \\
  \times & 0 & \times\\
  \Delta& \times & \times\\
\end{array}
\right)$& $\left(
\begin{array}{ccccccc}
    \Delta&\times  & \Delta \\
  \times & \times & 0\\
  \Delta& 0 & \times \\
\end{array}
\right)$& $\left(
\begin{array}{ccccccc}
    \Delta& \times & \Delta \\
  \times & \times & \times\\
  \Delta& \times & 0 \\
\end{array}
\right)$ \\ \hline

\hline 3 &$\left(
\begin{array}{ccccccc}
0 & \times & \Delta \\
\times & \times & \Delta\\
\Delta & \Delta& \times \\
\end{array}
\right)$ & $\left(
\begin{array}{ccccccc}
\times & 0 & \Delta \\
0 & \times & \Delta\\
\Delta& \Delta & \times \\
\end{array}
\right)$ & $\left(
\begin{array}{ccccccc}
    \Delta & \times & 0 \\
  \times & \times & \Delta\\
  0& \Delta & \times \\
\end{array}
\right)$& $\left(
\begin{array}{ccccccc}
    \Delta & \times & 0 \\
  \times & 0 & \Delta\\
  0& \Delta & \times \\
\end{array}
\right)$ & $\left(
\begin{array}{ccccccc}
    \Delta&\times  & \times \\
  \times & \Delta & 0\\
  \times& 0 & \times \\
\end{array}
\right)$& $\left(
\begin{array}{ccccccc}
    \Delta& \times & \times \\
  \times & \Delta & \times\\
  \times & \times & 0 \\
\end{array}
\right)$ \\ \hline

\hline 4 &$\left(
\begin{array}{ccccccc}
0 & \times & \times \\
\times & \Delta & \Delta\\
\times& \Delta& \times \\
\end{array}
\right)$ & $\left(
\begin{array}{ccccccc}
\times & 0 & \Delta \\
0 & \times & \times\\
\Delta & \times & \Delta \\
\end{array}
\right)$ & $\left(
\begin{array}{ccccccc}
    \Delta & \times & 0 \\
  \times & \times & \times\\
  0& \times & \Delta \\
\end{array}
\right)$ &  $\left(

\begin{array}{ccccccc}
    \Delta&\times  & \times \\
  \times & 0 & \times\\
  \times& \times & \Delta \\
\end{array}
\right)$&  
$\left(\begin{array}{ccccccc}
    \Delta&\times  & \times \\
  \times &\times & 0\\
  \times& 0 & \Delta \\
\end{array}
\right)$& $\left(
\begin{array}{ccccccc}
    \Delta& \times & \times \\
  \times & \times & \Delta\\
  \times & \Delta & 0 \\
\end{array}
\right)$ \\ \hline

\hline 5 &$\left(
\begin{array}{ccccccc}
0 & \times & \times \\
\times & \Delta & \times\\
\times& \times & \Delta \\
\end{array}
\right)$ & $\left(
\begin{array}{ccccccc}
\Delta & 0 & \times \\
0 & \Delta & \times\\
\times & \times & \times \\
\end{array}
\right)$ & $\left(
\begin{array}{ccccccc}
    \times & \Delta & 0 \\
  \Delta & \Delta & \times\\
  0& \times & \times \\
\end{array}
\right)$ & $\left(
\begin{array}{ccccccc}
    \times& \Delta &\Delta \\
  \Delta & 0 & \times\\
  \Delta & \times & \times\\
\end{array}
\right)$ & $\left(
\begin{array}{ccccccc}
    \times& \Delta &\Delta \\
  \Delta & \times & 0 \\
  \Delta & 0 & \times\\
\end{array}
\right)$& $\left(
\begin{array}{ccccccc}
    \times& \Delta & \Delta \\
  \Delta & \times & \times\\
  \Delta & \times & 0 \\
\end{array}
\right)$ \\ \hline

\hline 6 &$\left(
\begin{array}{ccccccc}
0 & \times & \times \\
\times & \times & \Delta\\
\times& \Delta& \Delta \\
\end{array}
\right)$ &$\left(
\begin{array}{ccccccc}
\Delta & 0 & \times \\
0 & \times & \Delta\\
\times & \Delta & \times \\
\end{array}
\right)$ & $\left(
\begin{array}{ccccccc}
    \times & \Delta & 0 \\
  \Delta & \times & \Delta\\
  0& \Delta & \times \\
\end{array}
\right)$ & $\left(
\begin{array}{ccccccc}
    \times& \Delta &\times \\
  \Delta & 0 & \Delta\\
  \times & \Delta & \times\\
\end{array}
\right)$ & $\left(
\begin{array}{ccccccc}
    \times& \Delta &\times \\
  \Delta & \Delta & 0\\
  \times & 0 & \times\\
\end{array}
\right)$&  $\left(
\begin{array}{ccccccc}
    \times& \Delta & \times \\
  \Delta & \Delta & \times\\
  \times & \times & 0 \\
\end{array}
\right)$ \\ \hline

\hline 7 &$\left(
\begin{array}{ccccccc}
0 & \Delta & \times \\
\Delta & \times & \Delta\\
\times& \Delta& \times \\
\end{array}
\right)$ &$\left(
\begin{array}{ccccccc}
\Delta & 0 & \times \\
0 & \times & \times\\
\times & \times & \Delta \\
\end{array}
\right)$ & $\left(
\begin{array}{ccccccc}
    \times & \Delta & 0 \\
  \Delta & \times & \times\\
  0& \times & \Delta \\
\end{array}
\right)$ & $\left(
\begin{array}{ccccccc}
    \times&\Delta  & \times \\
  \Delta & 0 & \times\\
  \times & \times & \Delta \\
\end{array}
\right)$ & $\left(
\begin{array}{ccccccc}
    \times&\Delta  & \times \\
  \Delta & \times & 0 \\
  \times & 0 & \Delta \\
\end{array}
\right)$& $\left(
\begin{array}{ccccccc}
    \times& \Delta & \times \\
  \Delta & \times & \Delta\\
  \times & \Delta & 0 \\
\end{array}
\right)$ \\ \hline

\hline 8 &$\left(
\begin{array}{ccccccc}
0 & \times & \Delta\\
\times & \times & \times\\
\Delta& \times& \Delta \\
\end{array}
\right)$ &$\left(
\begin{array}{ccccccc}
\times & 0 & \times \\
0 & \Delta & \Delta\\
\times & \Delta & \times \\
\end{array}
\right)$ & $\left(
\begin{array}{ccccccc}
    \times & \times & 0 \\
  \times & \Delta & \Delta\\
  0& \Delta & \times \\
\end{array}
\right)$ & $\left(
\begin{array}{ccccccc}
    \times&\times  & \Delta \\
  \times & 0 & \Delta\\
  \Delta & \Delta & \times \\
\end{array}
\right)$& $\left(
\begin{array}{ccccccc}
    \times&\times  & \Delta \\
  \times & \Delta & 0\\
  \Delta & 0 & \times \\
\end{array}
\right)$& $\left(
\begin{array}{ccccccc}
    \times& \times & \Delta \\
  \times & \Delta & \times\\
  \Delta & \times & 0 \\
\end{array}
\right)$ \\ \hline

\hline 9 &$\left(
\begin{array}{ccccccc}
0 & \times & \Delta \\
\times & \Delta & \times\\
\Delta& \times& \times \\
\end{array}
\right)$  & $\left(
\begin{array}{ccccccc}
    \times & 0 & \times \\
  0 & \Delta & \times\\
  \times & \times & \Delta \\
\end{array}
\right)$ &$\left(
\begin{array}{ccccccc}
    \times& \times &0 \\
  \times & \Delta & \times\\
  0 & \times & \Delta\\
\end{array}
\right)$ & $\left(
\begin{array}{ccccccc}
    \times&\times  & \Delta \\
  \times & 0 & \times\\
  \Delta & \times & \Delta \\
\end{array}
\right)$& $\left(
\begin{array}{ccccccc}
    \times&\times  & \Delta \\
  \times & \times & 0\\
  \Delta & 0 & \Delta \\
\end{array}
\right)$& $\left(
\begin{array}{ccccccc}
    \times& \times & \Delta \\
  \times & \times& \Delta\\
  \Delta & \Delta & 0 \\
\end{array}
\right)$ \\ \hline

\hline 10 &$\left(
\begin{array}{ccccccc}
0 & \Delta & \times \\
\Delta & \times & \times\\
\times& \times& \Delta \\
\end{array}
\right)$ &$\left(
\begin{array}{ccccccc}
\times & 0 & \times \\
0 & \times & \Delta\\
\times & \Delta & \Delta \\
\end{array}
\right)$ & $\left(
\begin{array}{ccccccc}
    \times & \times & 0 \\
  \times & \times & \Delta\\
  0& \Delta & \Delta \\
\end{array}
\right)$ &$\left(
\begin{array}{ccccccc}
    \times& \times &\times \\
  \times & 0 & \Delta\\
  \times & \Delta & \Delta\\
\end{array}
\right)$ & $\left(
\begin{array}{ccccccc}
    \times&\times  & \times \\
  \times & \Delta & 0\\
  \times & 0 & \Delta \\
\end{array}
\right)$&  $\left(
\begin{array}{ccccccc}
    \times& \times & \times \\
  \times & \Delta& \Delta\\
  \times & \Delta & 0 \\
\end{array}
\right)$ \\ \hline
\end{tabular}}
\caption{\label{tab2} Sixty phenomenologically possible hybrid texture structures of $M_{\nu}$ at 3$\sigma$ C.L where the triangles "$\bigtriangleup" $ denote equal and non-zero elements (or cofactors), and "$0$" denotes the vanishing element (or minor).$ "\times"$ denote the non-zero elements or cofactor.}
\end{footnotesize}
\end{center}
\end{table}
\newpage
\begin{table}[ht]
\begin{center}
\begin{footnotesize}
\resizebox{12cm}{!}{

\begin{tabular}{|c|c|c|c|c|}
%  % after \\: \hline or \cline{col1-col2} \cline{col3-col4} ...
  \hline
&\multicolumn{2}{c|}{1TEE} &\multicolumn{2}{c|}{1TEC} \\ 
\hline 
Cases&NO&IO&NO&IO\\
\hline
$A_{1}$  & $\rho=
-90^{0}-90^{0}$ & $\times$& $\times$
&$\rho=-0.0277^{0}--0.0220^{0} \oplus 0.0214^{0}-0.0274^{0}$  \\ 
& $\sigma= -90^{0}-90^{0}$ &
$\times$ & $\times $ & $\sigma=-0.0273^{0}-0.0271^{0}$ \\
&$\delta= 0^{0}-360^{0}$ & $\times$
& $\times$ & $\delta= 0^{0}-360^{0}$ \\ 
&$m_{1}=0.00147-0.0106$&$\times$&$\times$&$m_{1}=0.0430-0.0534$\\
&$m_{2}=0.00849-0.0139$&$\times$&$\times$&$m_{2}=0.0439-0.0541$\\
&$m_{3}=0.0437-0.0551$&$\times$&$\times$&$m_{3}=0.000904-0.00504$\\

\hline $A_{2}(A_{8})$  &
$\times$& $\times$ &
$\times$ &$\times$   \\
 \hline 
  $A_{3}(A_{7})$  &
$\times$& $\times$ &
$\times$ &$\times$   \\
 \hline 
$A_{4}(A_{6})$  &$\rho=
-90^{0}-90^{0}$ & $\times$& $\times$
&$\rho=-0.0275^{0}--0.0222^{0} \oplus 0.0215^{0}-0.0269^{0}$  \\ 
& $\sigma= -90^{0}-90^{0}$ &
$\times$ & $\times $ & $\sigma=-0.0250^{0}-0.0268^{0}$ \\
&$\delta= 0^{0}-360^{0}$ & $\times$
& $\times$ & $\delta= 0^{0}-360^{0}$ \\ 
&$m_{1}=0.00148-0.0106$&$\times$&$\times$&$m_{1}=0.0431-0.0534$\\
&$m_{2}=0.00850-0.0139$&$\times$&$\times$&$m_{2}=0.0439-0.0540$\\
&$m_{3}=0.0437-0.0551$&$\times$&$\times$&$m_{3}=0.000950-0.00504$\\
\hline
$A_{5}(A_{5})$   &$\rho=
-90^{0}-90^{0}$ & $\times$& $\times$
&$\times$  \\ 
& $\sigma= -90^{0}-90^{0}$ &
$\times$ & $\times $ & $\times$ \\
&$\delta= 0^{0}-360^{0}$ & $\times$
& $\times$ &$\times$ \\ 
&$m_{1}=0.00163-0.0105$&$\times$&$\times$&$\times$\\
&$m_{2}=0.00854-0.0137$&$\times$&$\times$&$\times$\\
&$m_{3}=0.0438-0.0545$&$\times$&$\times$&$\times$\\
\hline 
  $A_{9}(A_{10})$  &
$\times$& $\times$ &
$\times$ &$\times$   \\
 \hline 
 \end{tabular}}
\caption{\label{tab3}The allowed ranges of Dirac
CP-violating phase $\delta$, the  Majorana phases
$\rho, \sigma$,  three neutrino masses $m_{1},
m_{2}, m_{3}$ for the experimentally allowed
cases of Category A. Masses are in eV. $"\times"$ denotes the non-viability of case for a particular mass ordering.}
\end{footnotesize}
\end{center}
\end{table}

\begin{table}[ht]
\begin{center}
\begin{footnotesize}
\resizebox{14cm}{!}{
\begin{tabular}{|c|c|c|c|c|}
%  % after \\: \hline or \cline{col1-col2} \cline{col3-col4} ...
  \hline
&\multicolumn{2}{c|}{1TEE} &\multicolumn{2}{c|}{1TEC} \\ 
\hline 
Cases&NO&IO&NO&IO\\
\hline
$B_{1}(C_{1})$  & $\rho=
-90^{0}-90^{0}$ & $\times$& $\times$
&$\rho=-0.0277^{0}-0.0279^{0}$  \\ 
& $\sigma= -90^{0}-90^{0}$ &
$\times$ & $\times $ & $\sigma=-0.0284^{0}-0.0284^{0}$ \\
&$\delta= 0^{0}-360^{0}$ & $\times$
& $\times$ & $\delta= 0^{0}-360^{0}$ \\ 
&$m_{1}=0.00550-0.0298$&$\times$&$\times$&$m_{1}=0.0439-0.0597$\\
&$m_{2}=0.0101-0.0310$&$\times$&$\times$&$m_{2}=0.0447-0.0604$\\
&$m_{3}=0.0443-0.0605$&$\times$&$\times$&$m_{3}=0.000754-0.0297$\\

\hline
$B_{2}(C_{7})$  & $\rho=
-2.76^{0}-2.81^{0}$ & $\rho=-7.74^{0}-7.75^{0}$& $\rho=
-90^{0}-90^{0}$
&$\rho=-4.46^{0}-4.40^{0}$  \\ 
& $\sigma= -10.48^{0}-10.22^{0}$ &
$\sigma= -8.84^{0}-9.22^{0}$ & $\sigma=
-90^{0}-90^{0} $ & $\sigma=-5.41^{0}-5.39^{0}$ \\
&$\delta= 77.3^{0}-94.27^{0}\oplus 266.3^{0}-284^{0}$ & $\delta= 84.01^{0}-98.7^{0}\oplus 262.19^{0}-276.6^{0}$
& $\delta= 86.22^{0}-273.3^{0}$ & $\delta= 81.56^{0}-96.48^{0}\oplus 263.7^{0}-278.4^{0}$ \\ 
&$m_{1}=0.0235-0.314$&$m_{1}=0.0453-0.277$&$m_{1}=0.00548-0.267$&$m_{1}=0.0507-0.419$\\
&$m_{2}=0.0242-0.0310$&$m_{2}=0.0465-0.274$&$m_{2}=0.00806-0.316$&$m_{2}=0.0508-0.419$\\
&$m_{3}=0.0485-0.0315$&$m_{3}=0.0117-0.271$&$m_{3}=0.0428-0.320$&$m_{3}=0.0249-0.414$\\
 \hline 
 
$B_{3}(C_{6})$  & $\rho=
-90^{0}--5.9^{0} \oplus 5.9^{0}-90^{0}$ & $\rho=-90^{0}-90^{0}$& $\rho=
-90^{0}-90^{0}$
&$\rho=-90^{0}-90^{0}$  \\ 
& $\sigma=
-90^{0}--5.9^{0} \oplus 5.9^{0}-90^{0}$ &
$\sigma= -90^{0}-90^{0}$ & $\sigma=
-90^{0}-90^{0} $ & $\sigma=-90^{0}-90^{0}$ \\
&$\delta= 7.47^{0}-356^{0}$ & $\delta= 8.3^{0}-173^{0}\oplus 187^{0}-351^{0}$
& $\delta= 5.16^{0}-178^{0}\oplus 183.7^{0}-353.6^{0}$ & $\delta= 0^{0}-360^{0}$ \\ 
&$m_{1}=0.0153-0.331$&$m_{1}=0.0459-0.335$&$m_{1}=0.00504-0.485$&$m_{1}=0.0427-0.505$\\
&$m_{2}=0.0161-0.331$&$m_{2}=0.0461-0.334$&$m_{2}=0.00920-0.483$&$m_{2}=0.0445-0.503$\\
&$m_{3}=0.0445-0.0335$&$m_{3}=0.0136-0.330$&$m_{3}=0.0416-0.480$&$m_{3}=0.00884-0.500$\\
 \hline 
$B_{4}(C_{5})$  & $\rho=
-19.07^{0}-19.75^{0}$ & $\rho=-4.80^{0}-4.66^{0}$& $\rho=
-43.13^{0}-40.12^{0}$
&$\rho=-16.38^{0}-18.8^{0}$  \\ 
& $\sigma= -90^{0}-90^{0}$ &
$\sigma= -2.83^{0}-3.12^{0}$ & $\sigma=
-43^{0}-40.1^{0} $ & $\sigma=-41.43^{0}-40.59^{0}$ \\
&$\delta= 0^{0}-100^{0}\oplus 260.2^{0}-360^{0}$ & $\delta= 84.68^{0}-94.7^{0}\oplus 265.08^{0}-275.5^{0}$
& $\delta= 88.04^{0}-124.4^{0}\oplus 230.09^{0}-272^{0}$ & $\delta= 5.56^{0}-13.82^{0}\oplus 255.7^{0}-307.3^{0}$ \\ 
&$m_{1}=0.00681-0.318$&$m_{1}=0.0488-0.314$&$m_{1}=0.0105-0.373$&$m_{1}=0.0444-0.420$\\
&$m_{2}=0.00969-0.0317$&$m_{2}=0.0486-0.312$&$m_{2}=0.0129-0.369$&$m_{2}=0.0447-0.416$\\
&$m_{3}=0.0421-0.0321$&$m_{3}=0.0202-0.310$&$m_{3}=0.0435-0.374$&$m_{3}=0.0103-0.418$\\
 \hline 
 $B_{5}(C_{4})$  & $\rho=
-20.83^{0}-20.79^{0}$ & $\rho=-90^{0}--11.3^{0}\oplus 11.3^{0}-90^{0}$& $\rho=
-90^{0}-90^{0}$
&$\rho=-24.19^{0}-24.45^{0}$  \\ 
& $\sigma= -36.3^{0}-36^{0}$ &
$\$\sigma=-90^{0}--14.3^{0}\oplus 14.8^{0}-90^{0}$ & $\sigma=
-90^{0}-90^{0} $ & $\sigma=-27.7^{0}-27.4^{0}$ \\
&$\delta= 0^{0}-177.5^{0}\oplus 182^{0}-360^{0}$ & $\delta= 0^{0}-168^{0}\oplus 192.5^{0}-360^{0}$
& $\delta= 9.33^{0}-172.4^{0}\oplus 187.2{0}-350.4^{0}$ & $\delta= 0^{0}-179.3^{0}\oplus 180.6^{0}-360^{0}$ \\ 
&$m_{1}=0.00965-0.330$&$m_{1}=0.0761-0.330$&$m_{1}=0.0603-0.498$&$m_{1}=0.0440-0.498$\\
&$m_{2}=0.0113-0.0330$&$m_{2}=0.0762-0.330$&$m_{2}=0.0608-0.496$&$m_{2}=0.0447-0.499$\\
&$m_{3}=0.0421-0.0332$&$m_{3}=0.0603-0.328$&$m_{3}=0.0629-0.500$&$m_{3}=0.0103-0.499$\\
\hline
 $B_{6}(C_{3})$ & $\rho=
-90^{0}-90^{0}$ & $\times$& $\times$
&$\rho=-90^{0}-90^{0}$  \\ 
& $\sigma= -90^{0}-90^{0}$ &
$\times$ & $\times $ & $\sigma=-90^{0}-90^{0}$ \\
&$\delta= 0^{0}-360^{0}$ & $\times$
& $\times$ & $\delta= 0^{0}-360^{0}$ \\ 
&$m_{1}=0.0147-0.324$&$\times$&$\times$&$m_{1}=0.0452-0.454$\\
&$m_{2}=0.00161-0.324$&$\times$&$\times$&$m_{2}=0.0451-0.451$\\
&$m_{3}=0.0445-0.324$&$\times$&$\times$&$m_{3}=0.00161-0.450$\\

\hline
 
$B_{7}(C_{2})$ & $\rho=
-26.25^{0}-26.43^{0}$ & $\times$& $\times$
&$\rho=-30.92^{0}-30.42^{0}$  \\ 
& $\sigma= -50.28^{0}-50.2^{0}$ &
$\times$ & $\times $ & $\sigma=-41.9^{0}-43.8^{0}$ \\
&$\delta=  0^{0}-177.5^{0}\oplus 182^{0}-360^{0}$ & $\times$
& $\times$ & $\delta= 0^{0}-360^{0}$ \\ 
&$m_{1}=0.00883-0.331$&$\times$&$\times$&$m_{1}=0.0447-0.498$\\
&$m_{2}=0.00121-0.329$&$\times$&$\times$&$m_{2}=0.0451-0.498$\\
&$m_{3}=0.0421-0.334$&$\times$&$\times$&$m_{3}=0.00804-0.498$\\

\hline
$B_{8}(C_{10})$ & $\rho=
-90^{0}-90^{0}$ & $\rho=
-90^{0}-90^{0}$& $\rho=
-90^{0}-90^{0}$
&$\rho=-90^{0}-90^{0}$  \\ 
& $\sigma= -90^{0}-90^{0}$ &
$\sigma= -90^{0}-90^{0}$ & $\sigma= -90^{0}-90^{0}$ & $\sigma=-90^{0}-90^{0}$ \\
&$\delta=  0^{0}-166.5^{0}\oplus 196.5^{0}-360^{0}$ & $\delta=  45.05^{0}-135^{0}\oplus 224.5^{0}-314.6^{0}$
& $\delta= 45.86^{0}-315.05^{0}$ & $\delta= 0^{0}-360^{0}$ \\ 
&$m_{1}=0.00127-0.279$&$m_{1}=0.0418-0.279$&$m_{1}=0.00498-0.381$&$m_{1}=0.0427-0.303$\\
&$m_{2}=0.00691-0.277$&$m_{2}=0.0429-0.307$&$m_{2}=0.00736-0.382$&$m_{2}=0.0428-0.302$\\
&$m_{3}=0.0409-0.280$&$m_{3}=0.00398-0.305$&$m_{3}=0.0407-0.384$&$m_{3}=0.00342-0.298$\\
\hline
$B_{9}(C_{9})$ & $\rho=
-90^{0}-90^{0}$ & $\rho=
-90^{0}-90^{0}$& $\rho=
-90^{0}-90^{0}$
&$\rho=-90^{0}-90^{0}$  \\ 
& $\sigma= -90^{0}-90^{0}$ &
$\sigma= -90^{0}-90^{0}$ & $\sigma= -90^{0}-90^{0}$ & $\sigma=-90^{0}-90^{0}$ \\
&$\delta=  5.55^{0}-354^{0}$ & $\delta=  12^{0}-171.7^{0}\oplus 189.7^{0}-352.7^{0}$
&$\delta=8.32^{0}-175.9^{0}\oplus 186^{0}-350.47^{0}$  & $\delta= 0^{0}-360^{0}$ \\ 
&$m_{1}=0.00338-0.279$&$m_{1}=0.0435-0.319$&$m_{1}=0.00622-0.487$&$m_{1}=0.0447-0.486$\\
&$m_{2}=0.00831-0.277$&$m_{2}=0.0445-0.319$&$m_{2}=0.00681-0.486$&$m_{2}=0.0451-0.487$\\
&$m_{3}=0.0423-0.282$&$m_{3}=0.00398-0.317$&$m_{3}=0.00914-0.486$&$m_{3}=0.00458-0.480$\\
\hline 
$B_{10}(C_{8})$ & $\rho=
-90^{0}-90^{0}$ & $\rho=
-90^{0}-90^{0}$& $\rho=
-90^{0}-90^{0}$
&$\rho=-90^{0}-90^{0}$  \\ 
& $\sigma= -90^{0}-90^{0}$ &
$\sigma= -90^{0}-90^{0}$ & $\sigma= -90^{0}-90^{0}$ & $\sigma=-90^{0}-90^{0}$ \\
&$\delta=  0^{0}-360^{0}$ & $\delta=  44.64^{0}-136.5^{0}\oplus 224.6^{0}-314.2^{0}$
&$\delta=1.92^{0}-313.04^{0}$  & $\delta= 0^{0}-360^{0}$ \\ 
&$m_{1}=0.00163-0.294$&$m_{1}=0.0425-0.277$&$m_{1}=0.00516-0.348$&$m_{1}=0.0427-0.365$\\
&$m_{2}=0.00691-0.293$&$m_{2}=0.0423-0.275$&$m_{2}=0.00806-0.346$&$m_{2}=0.0453-0.365$\\
&$m_{3}=0.0409-0.299$&$m_{3}=0.00380-0.272$&$m_{3}=0.0416-0.348$&$m_{3}=0.00879-0.391$\\
\hline
\end{tabular}}
\caption{\label{tab4}The allowed ranges of Dirac
CP-violating phase $\delta$, the  Majorana phases
$\rho, \sigma$,  three neutrino masses $m_{1},
m_{2}, m_{3}$ for the experimentally allowed
cases of Category B(C). Masses are in eV. $"\times"$ denotes the non-viability of case for a particular mass ordering.}
\end{footnotesize}
\end{center}
\end{table}

\begin{table}[ht]
\begin{center}
\begin{footnotesize}
\resizebox{14cm}{!}{
\begin{tabular}{|c|c|c|c|c|}
%  % after \\: \hline or \cline{col1-col2} \cline{col3-col4} ...
  \hline
&\multicolumn{2}{c|}{1TEE} &\multicolumn{2}{c|}{1TEC} \\ 
\hline 
Cases&NO&IO&NO&IO\\
\hline
$D_{1}(F_{2})$  & $\rho=
-90^{0}-90^{0}$ & $\rho=
-90^{0}-90^{0}$& $\rho=
-90^{0}-90^{0}$
&$\rho=
-90^{0}-90^{0}$  \\ 
& $\sigma= -90^{0}-90^{0}$ &
$\sigma= -90^{0}-90^{0}$ & $\sigma= -90^{0}-90^{0}$ & $\sigma= -90^{0}-90^{0}$ \\
&$\delta= 30.185^{0}-322.6^{0}$ & 
$\delta= 0^{0}-360^{0}$ & $\delta= 0^{0}-360^{0}$&$\delta= 31.79^{0}-318.66^{0}$ \\ 
&$m_{1}=0.0541-0.224$&$m_{1}=0.00432-0.252$&$m_{1}=0.00331-0.603$&$m_{1}=0.0731-0.501$\\
&$m_{2}=0.0538-0.224$&$m_{2}=0.00437-0.251$&$m_{2}=0.00800-0.606$&$m_{2}=0.0719-0.500$\\
&$m_{3}=0.0679-0.230$&$m_{3}=0.00343-0.248$&$m_{3}=0.0371-0.604$&$m_{3}=0.0593-0.496$\\
\hline
$D_{2}(F_{1})$  & $\rho=-50.28^{0}-60.48^{0}$ & $\rho=-90^{0}-90^{0}$& $\rho=
-90^{0}-90^{0}$
&$\rho=-63.3^{0}-64^{0}$  \\ 
& $\sigma=
-90^{0}-90^{0}$&$\sigma=
-90^{0}-90^{0} $  & $\sigma=
-90^{0}-90^{0} $ & $\sigma=
-89.37^{0}-86.25^{0}$ \\
&$\delta= 87.19^{0}-297.4^{0}$ &$\delta= 0^{0}-360^{0}$ & $\delta= 0^{0}-64.34^{0}\oplus 295.7^{0}-360^{0}$
& $\delta= 70.50^{0}-295.77^{0}$ \\ 
&$m_{1}=0.124-0.293$&$m_{1}=0.0422-0.263$&$m_{1}=0.00542-0.0277$&$m_{1}=0.128-0.581$\\
&$m_{2}=0.125-0.294$&$m_{2}=0.0430-0.262$&$m_{2}=0.00993-0.0291$&$m_{2}=0.130-0.582$\\
&$m_{3}=0.132-0.295$&$m_{3}=0.00552-0.257$&$m_{3}=0.0440-0.0600$&$m_{3}=0.120-0.579$\\
 \hline 
$D_{3}(F_{4})$  & $\times$ & $\rho=-90^{0}-90^{0}$& $\rho=
-90^{0}-90^{0}$
&$\times$  \\ 
& $\times$&$\sigma=
-90^{0}-90^{0} $  & $\sigma=
-90^{0}-90^{0} $ & $\times$ \\
&$\times$ &$\delta= 0^{0}-360^{0}$ & $\delta= 0^{0}-64.34^{0}\oplus 295.7^{0}-360^{0}$
& $\times$ \\ 
&$\times$&$m_{1}=0.0439-0.308$&$m_{1}=0.0161-0.474$&$\times$\\
&$\times$&$m_{2}=0.0445-0.308$&$m_{2}=0.0164-0.471$&$\times$\\
&$\times$&$m_{3}=0.00888-0.304$&$m_{3}=0.0416-0.469$&$\times$\\
 \hline 
$D_{4}(F_{3})$  & $\rho=
-58.71^{0}-59.35^{0}$ & $\rho=
-90^{0}-90^{0}$& $\rho=
-90^{0}-90^{0}$
&$\rho=
-61.4^{0}-61.5^{0}$  \\ 
& $\sigma= -88.90^{0}-87.18^{0}$ &
$\sigma= -90^{0}-90^{0}$ & $\sigma= -90^{0}-90^{0}$ & $\sigma= -90^{0}-90^{0}$ \\
&$\delta= 75.87^{0}-285.58^{0}$ & 
 $\delta= 0^{0}-360^{0}$ & $\delta= 18.12^{0}-318.75^{0}$&$\delta= 77.86^{0}-286.15^{0}$ \\ 
&$m_{1}=0.0574-0.287$&$m_{1}=0.00429-0.254$&$m_{1}=0.0102-0.795$&$m_{1}=0.0711-0.424$\\
&$m_{2}=0.0575-0.286$&$m_{2}=0.00430-0.254$&$m_{2}=0.0147-0.796$&$m_{2}=0.0740-0.424$\\
&$m_{3}=0.0679-0.291$&$m_{3}=0.00969-0.202$&$m_{3}=0.0369-0.790$&$m_{3}=0.0564-0.418$\\
\hline
 $D_{5}(F_{5})$  & $\rho=
-90^{0}-90^{0}$ & $\rho=-90^{0}-90^{0}$& $\rho=
-90^{0}-90^{0}$
&$\rho=-90^{0}-90^{0}$  \\ 
& $\sigma=
-90^{0}-90^{0}$ &
$\sigma= -90^{0}-90^{0}$ & $\sigma=
-90^{0}-90^{0} $ & $\sigma=-90^{0}-90^{0}$ \\
&$\delta= 0^{0}-360^{0}$ & $\delta= 0^{0}-360^{0}$
& $\delta= 0^{0}-360^{0}$ & $\delta= 0^{0}-360^{0}$ \\ 
&$m_{1}=0.0337-0.334$&$m_{1}=0.0431-0.335$&$m_{1}=0.00255-0.509$&$m_{1}=0.0538-0.505$\\
&$m_{2}=0.0338-0.334$&$m_{2}=0.0434-0.334$&$m_{2}=0.00370-0.505$&$m_{2}=0.0539-0.506$\\
&$m_{3}=0.0541-0.334$&$m_{3}=0.00940-0.330$&$m_{3}=0.0351-0.510$&$m_{3}=0.0305-0.500$\\
 \hline 
$D_{6}(F_{9})$  & $\rho=
-65.09^{0}-58.48^{0}$ & $\rho=-90^{0}-90^{0}$ & $\rho=
-23.03^{0}--7.74^{0} \oplus 7.66^{0}-22.14^{0}$
&$\rho=-18.42^{0}--6.56^{0} \oplus 6.77^{0}-18.27^{0}$  \\ 
& $\sigma= -90^{0}-90^{0}$ &
$\sigma= -90^{0}-90^{0}$ & $\sigma=
-68.53^{0}--38.75^{0} \oplus 37.86^{0}-69.0^{0}$ & $\sigma=
-71.1^{0}--46.2^{0} \oplus 46.2^{0}-70.48^{0}$ \\
&$\delta= 64.2^{0}-288.1^{0}$ & $\delta= 0^{0}-360^{0}$
& $\delta= 109.68^{0}-150.05^{0} \oplus 210.4^{0}-250.2^{0}$ & $\delta= 118.19^{0}-154.2^{0}\oplus 206.7^{0}-242.3^{0}$ \\ 
&$m_{1}=0.0974-0.323$& $m_{1}=0.0446-0.277$& $m_{1}=0.0202-0.456$& $m_{1}=0.108-0.398$\\
&$m_{2}=0.0965-0.0321$& $m_{2}=0.0448-0.277$& $m_{2}=0.0204-0.456$& $m_{2}=0.109-0.396$\\
&$m_{3}=0.107-0.0326$& $m_{3}=0.00888-0.324$& $m_{3}=0.0451-0.456$& $m_{3}=0.0962-0.396$\\
 \hline
 $D_{7}(F_{8})$  & $\rho=
-90^{0}-90^{0}$ & $\rho=-90^{0}-90^{0}$& $\rho=
-90^{0}-90^{0}$
&$\rho=-90^{0}-90^{0}$  \\ 
& $\sigma=
-90^{0}-90^{0}$ &
$\sigma= -90^{0}-90^{0}$ & $\sigma=
-90^{0}-90^{0} $ & $\sigma=-90^{0}-90^{0}$ \\
&$\delta= 0^{0}-360^{0}$ & $\delta= 0^{0}-360^{0}$
& $\delta= 0^{0}-360^{0}$ & $\delta= 0^{0}-360^{0}$ \\ 
&$m_{1}=0.0478-0.273$&$m_{1}=0.0439-0.333$&$m_{1}=0.00663-0.436$&$m_{1}=0.0647-0.437$\\
&$m_{2}=0.0479-0.273$&$m_{2}=0.0454-0.333$&$m_{2}=0.0103-0.434$&$m_{2}=0.0649-0.438$\\
&$m_{3}=0.0644-0.300$&$m_{3}=0.00806-0.330$&$m_{3}=0.0416-0.438$&$m_{3}=0.0439-0.491$\\
 \hline 
$D_{8}(F_{7})$  & $\times$ & $\rho=-90^{0}-90^{0}$& $\rho=
-90^{0}-90^{0}$
&$\times$  \\ 
& $\times$&$\sigma=
-90^{0}-90^{0} $  & $\sigma=
-90^{0}-90^{0} $ & $\times$ \\
&$\times$ &$\delta= 0^{0}-360^{0}$ & $\delta= 0^{0}-64.34^{0}\oplus 295.7^{0}-360^{0}$
& $\times$ \\ 
&$\times$&$m_{1}=0.0451-0.0938$&$m_{1}=0.00941-0.0803$&$\times$\\
&$\times$&$m_{2}=0.0459-0.0941$&$m_{2}=0.0125-0.0804$&$\times$\\
&$\times$&$m_{3}=0.00121-0.0790$&$m_{3}=0.0445-0.0953$&$\times$\\
 \hline
 $D_{9}(F_{6})$  & $\rho=
-90^{0}-90^{0}$ & $\rho=-90^{0}-90^{0}$& $\rho=
-90^{0}-90^{0}$
&$\rho=-90^{0}-90^{0}$  \\ 
& $\sigma=
-90^{0}-90^{0}$ &
$\sigma= -90^{0}-90^{0}$ & $\sigma=
-90^{0}-90^{0} $ & $\sigma=-90^{0}-90^{0}$ \\
&$\delta= 0^{0}-360^{0}$ & $\delta= 0^{0}-360^{0}$
& $\delta= 0^{0}-360^{0}$ & $\delta= 0^{0}-360^{0}$ \\ 
&$m_{1}=0.0492-0.300$&$m_{1}=0.0455-0.310$&$m_{1}=0.00769-0.436$&$m_{1}=0.0647-0.411$\\
&$m_{2}=0.0496-0.300$&$m_{2}=0.0456-0.310$&$m_{2}=0.00932-0.434$&$m_{2}=0.0638-0.412$\\
&$m_{3}=0.0638-0.303$&$m_{3}=0.00969-0.306$&$m_{3}=0.0426-0.438$&$m_{3}=0.0458-0.407$\\
 \hline   
$D_{10}(F_{10})$  & $\times$ & $\rho=-90^{0}-90^{0}$& $\rho=
-90^{0}-90^{0}$
&$\times$  \\ 
& $\times$&$\sigma=
-90^{0}-90^{0} $  & $\sigma=
-90^{0}-90^{0} $ & $\times$ \\
&$\times$ &$\delta= 0^{0}-360^{0}$ & $\delta= 0^{0}-64.34^{0}\oplus 295.7^{0}-360^{0}$
& $\times$ \\ 
&$\times$&$m_{1}=0.0432-0.102$&$m_{1}=0.00201-0.0793$&$\times$\\
&$\times$&$m_{2}=0.0442-0.102$&$m_{2}=0.00870-0.0795$&$\times$\\
&$\times$&$m_{3}=0.00301-0.0901$&$m_{3}=0.0435-0.0932$&$\times$\\
 \hline
\end{tabular}}
\caption{\label{tab5}The allowed ranges of Dirac
CP-violating phase $\delta$, the  Majorana phases
$\rho, \sigma$,  three neutrino masses $m_{1},
m_{2}, m_{3}$ for the experimentally allowed
cases of Category D(F). Masses are in eV. $"\times"$ denotes the non-viability of case for a particular mass ordering.}
\end{footnotesize}
\end{center}
\end{table}

\begin{table}[ht]
\begin{center}
\begin{footnotesize}
\resizebox{14cm}{!}{
\begin{tabular}{|c|c|c|c|c|}
%  % after \\: \hline or \cline{col1-col2} \cline{col3-col4} ...
  \hline
&\multicolumn{2}{c|}{1TEE} &\multicolumn{2}{c|}{1TEC} \\ 
\hline 
Cases&NO&IO&NO&IO\\
 \hline
$E_{1}(E_{2})$  & $\times$ & $\rho=-90^{0}-90^{0}$& $\rho=
-90^{0}-90^{0}$
&$\times$  \\ 
& $\times$&$\sigma=
-90^{0}-90^{0} $  & $\sigma=
-90^{0}-90^{0} $ & $\times$ \\
&$\times$ &$\delta= 0^{0}-360^{0}$ & $\delta= 0^{0}-64.34^{0}\oplus 295.7^{0}-360^{0}$
& $\times$ \\ 
&$\times$&$m_{1}=0.0511-0.0804$&$m_{1}=0.00283-0.0629$&$\times$\\
&$\times$&$m_{2}=0.0516-0.0805$&$m_{2}=0.00878-0.0634$&$\times$\\
&$\times$&$m_{3}=0.0264-0.0678$&$m_{3}=0.0440-0.0809$&$\times$\\
 \hline
$E_{3}(E_{4})$  & $\times$ & $\rho=-90^{0}-90^{0}$& $\rho=
-90^{0}-90^{0}$
&$\times$  \\ 
& $\times$&$\sigma=
-90^{0}-90^{0}$  & $\sigma=
-90^{0}-90^{0}$ & $\times$ \\
&$\times$ &$\delta= 0^{0}-360^{0}$ & $\delta= 0^{0}-64.34^{0}\oplus 295.7^{0}-360^{0}$
& $\times$ \\ 
&$\times$&$m_{1}=0.0439-0.333$&$m_{1}=0.00914-0.499$&$\times$\\
&$\times$&$m_{2}=0.0457-0.333$&$m_{2}=0.00115-0.504$&$\times$\\
&$\times$&$m_{3}=0.00888-0.333$&$m_{3}=0.0432-0.500$&$\times$\\
 \hline 
 $E_{5}$  & $\rho=-90^{0}--156 \bigoplus 29.3-90^{0}$ & $\rho=-90^{0}-90^{0}$& $\times$
&$\times$  \\ 
& $\rho=-90^{0}--156 \bigoplus 29.3-90^{0}$&$\sigma=
-90^{0}-90^{0} $  & $\times $ & $\times$ \\
&$\delta=29.3^{0}-153.5^{0}\oplus 206.5^{0}-334^{0}$ &$\delta= 0^{0}-360^{0}$ & $\times$
& $\times$ \\ 
&$m_{1}=0.130-0.332$&$m_{1}=0.0439-0.269$&$\times$&$\times$\\
&$m_{2}=0.132-0.330$&$m_{2}=0.0451-0.270$&$\times$&$\times$\\
&$m_{3}=0.137-0.334$&$m_{3}=0.0103-0.262$&$\times$&$\times$\\
 \hline 
$E_{6}(E_{9})$  & $\times$ & $\rho=-90^{0}-90^{0}$& $\rho=
-90^{0}-90^{0}$
&$\times$  \\ 
& $\times$&$\sigma=
-90^{0}-90^{0} $  & $\sigma=
-90^{0}-90^{0} $ & $\times$ \\
&$\times$ &$\delta= 0^{0}-360^{0}$ & $\delta= 0^{0}-360^{0}$
& $\times$ \\ 
&$\times$&$m_{1}=0.0511-0.333$&$m_{1}=0.00914-0.499$&$\times$\\
&$\times$&$m_{2}=0.0517-0.333$&$m_{2}=0.00115-0.504$&$\times$\\
&$\times$&$m_{3}=0.0273-0.333$&$m_{3}=0.0432-0.500$&$\times$\\
 \hline 
$E_{7}(E_{8})$  & $\times$ & $\rho=-90^{0}-90^{0}$& $\rho=
-90^{0}-90^{0}$
&$\times$  \\ 
& $\times$&$\sigma=
-90^{0}-90^{0} $  & $\sigma=
-90^{0}-90^{0} $ & $\times$ \\
&$\times$ &$\delta= 0^{0}-360^{0}$ & $\delta= 0^{0}-360^{0}$
& $\times$ \\ 
&$\times$&$m_{1}=0.0511-0.0908$&$m_{1}=0.0225-0.0469$&$\times$\\
&$\times$&$m_{2}=0.0517-0.0911$&$m_{2}=0.0238-0.0480$&$\times$\\
&$\times$&$m_{3}=0.0273-0.0746$&$m_{3}=0.0494-0.0708$&$\times$\\
 \hline
$E_{10}$  & $\rho=
-90^{0}--28.6^{0} \oplus 26.01^{0}-90^{0}$ & $\rho=-18.15^{0}-18.08^{0}$ & $\rho=
-90^{0}-90^{0}$
&$\rho=-90^{0}--19^{0} \oplus 19.5^{0}-90^{0}$  \\ 
&$\sigma=-90^{0}--28.6^{0} \oplus 26.01^{0}-90^{0}$
& $\sigma= -90^{0}--6.41^{0} \oplus 6.09^{0}-90^{0}$ &
$\sigma= -90^{0}-90^{0}$ & $\sigma=
-90^{0}--19^{0} \oplus 19.5^{0}-90^{0}$  \\
&$\delta= 25.35^{0}-151.92^{0} \oplus 206.97^{0}-332.3^{0}$ & $\delta= 0^{0}-177^{0} \oplus 184.3^{0}-357.5^{0}$
 & $\delta= 0^{0}-360^{0}$ & $\delta= 23.75^{0}-161.97^{0} \oplus 198.9^{0}-339^{0} $  \\ 
&$m_{1}=0.136-0.331$& $m_{1}=0.0431-0.334$& $m_{1}=0.00156-0.481$& $m_{1}=0.141-0.481$\\
&$m_{2}=0.137-0.331$& $m_{2}=0.0445-0.334$& $m_{2}=0.00571-0.480$& $m_{2}=0.142-0.492$\\
&$m_{3}=0.143-0.334$& $m_{3}=0.00969-0.330$& $m_{3}=0.0404-0.484$& $m_{3}=0.133-0.491$\\
\hline 
\end{tabular}}
\caption{\label{tab6} The allowed ranges of Dirac
CP-violating phase $\delta$, the  Majorana phases
$\rho, \sigma$,  three neutrino masses $m_{1},
m_{2}, m_{3}$ for the experimentally allowed
cases of Category E. Masses are in eV. $"\times"$ denotes the non-viability of case for a particular mass ordering.}
\end{footnotesize}
\end{center}
\end{table}

\end{document}